\pdfoutput=1
\documentclass[aip, rsi, reprint, nofootinbib]{revtex4-1}
\usepackage{graphicx, amsmath, amssymb}
\usepackage[usenames, dvipsnames]{color}
\usepackage{gensymb}
\usepackage{dcolumn}
\usepackage{bm}
\usepackage[utf8]{inputenc}
\usepackage[T1]{fontenc}
\usepackage{mathptmx}
\usepackage{multirow}

\newcommand{\pb}{\textsc{Polarbear}}

\begin{document}

\title{The POLARBEAR Fourier Transform Spectrometer Calibrator and Spectroscopic Characterization of the POLARBEAR Instrument}
\date{\today \\ 
\ \ This article may be downloaded for personal use only. \\
\ \ Any other use requires prior permission of the author and AIP Publishing. \\
\ \ This article appeared in Review of Scientific Instruments 90, 115115 (2019). \\
\ \ And may be found at https://doi.org/10.1063/1.5095160.}

\author{F. Matsuda}
\email{frederick.matsuda@ipmu.jp}
\affiliation{Kavli Institute for the Physics and Mathematics of the Universe (WPI), The University of Tokyo Institutes for Advanced Study, The University of Tokyo, Kashiwa, Chiba 277-8583, Japan}
\author{L. Lowry}
\affiliation{Department of Physics, University of California, San Diego, CA 92093-0424, USA}
\author{A. Suzuki}
\affiliation{Physics Division, Lawrence Berkeley National Laboratory, Berkeley, CA 94720, USA}
\author{M. Aguilar F\'aundez}
\affiliation{Department of Physics and Astronomy, Johns Hopkins University, Baltimore, MD 21218, USA}
\affiliation{Departamento de F\'isica, FCFM, Universidad de Chile, Blanco Encalada 2008, Santiago, Chile}
\author{K. Arnold}
\affiliation{Department of Physics, University of California, San Diego, CA 92093-0424, USA}
\author{D. Barron}
\affiliation{Department of Physics and Astronomy, University of New Mexico, Albuquerque, NM 87131, USA}
\author{F. Bianchini}
\affiliation{School of Physics, University of Melbourne, Parkville, VIC 3010, Australia}
\author{K. Cheung}
\affiliation{Department of Physics, University of California, Berkeley, CA 94720, USA}
\author{Y. Chinone}
\affiliation{Department of Physics, University of California, Berkeley, CA 94720, USA}
\affiliation{Kavli Institute for the Physics and Mathematics of the Universe (WPI), The University of Tokyo Institutes for Advanced Study, The University of Tokyo, Kashiwa, Chiba 277-8583, Japan}
\author{T. Elleflot}
\affiliation{Department of Physics, University of California, San Diego, CA 92093-0424, USA}
\author{G. Fabbian}
\affiliation{Department of Physics \& Astronomy, University of Sussex, Brighton BN1 9QH, UK}
\author{N. Goeckner-Wald}
\affiliation{Department of Physics, University of California, Berkeley, CA 94720, USA}
\author{M. Hasegawa}
\affiliation{High Energy Accelerator Research Organization (KEK), Tsukuba, Ibaraki 305-0801, Japan}
\author{D. Kaneko}
\author{N. Katayama}
\affiliation{Kavli Institute for the Physics and Mathematics of the Universe (WPI), The University of Tokyo Institutes for Advanced Study, The University of Tokyo, Kashiwa, Chiba 277-8583, Japan}
\author{B. Keating}
\affiliation{Department of Physics, University of California, San Diego, CA 92093-0424, USA}
\author{A.T. Lee}
\affiliation{Department of Physics, University of California, Berkeley, CA 94720, USA}
\affiliation{Physics Division, Lawrence Berkeley National Laboratory, Berkeley, CA 94720, USA}
\author{M. Navaroli}
\affiliation{Department of Physics, University of California, San Diego, CA 92093-0424, USA}
\author{H. Nishino}
\affiliation{High Energy Accelerator Research Organization (KEK), Tsukuba, Ibaraki 305-0801, Japan}
\author{H. Paar}
\affiliation{Department of Physics, University of California, San Diego, CA 92093-0424, USA}
\author{G. Puglisi}
\affiliation{Department of Physics, Stanford University, Stanford, CA, 94305}
\author{P.L. Richards}
\affiliation{Department of Physics, University of California, Berkeley, CA 94720, USA}
\author{J. Seibert}
\affiliation{Department of Physics, University of California, San Diego, CA 92093-0424, USA}
\author{P. Siritanasak}
\affiliation{Department of Physics, University of California, San Diego, CA 92093-0424, USA}
\affiliation{National Astronomical Research Institute of Thailand , Chiangmai, Thailand}
\author{O. Tajima}
\affiliation{Department of Physics, Kyoto University, Kyoto 606-8502, Japan}
\author{S. Takatori}
\affiliation{SOKENDAI (The Graduate University for Advanced Studies), Shonan Village, Hayama, Kanagawa 240-0193, Japan}
\affiliation{High Energy Accelerator Research Organization (KEK), Tsukuba, Ibaraki 305-0801, Japan}
\author{C. Tsai}
\affiliation{Department of Physics, University of California, San Diego, CA 92093-0424, USA}
\author{B. Westbrook}
\affiliation{Radio Astronomy Laboratory, University of California, Berkeley, CA 94720, USA}

\providecommand{\e}[1]{\ensuremath{\times 10^{#1}}}

\begin{abstract}
We describe the Fourier Transform Spectrometer (FTS) used for in-field testing of the \pb{} receiver, an experiment located in the Atacama Desert of Chile which measures the cosmic microwave background (CMB) polarization. The \pb{}-FTS (PB-FTS) is a Martin-Puplett interferometer designed to couple to the Huan Tran Telescope (HTT) on which the \pb{} receiver is installed. 
The PB-FTS measured the spectral response of the \pb{} receiver with signal-to-noise ratio (SNR) $>20$ for $\sim$69\% of the focal plane detectors due to three features: a high throughput of 15.1 steradian cm$^{2}$, optimized optical coupling to the \pb{} optics using a custom designed output parabolic mirror, and a continuously modulated output polarizer.
The PB-FTS parabolic mirror is designed to mimic the shape of the 2.5 m-diameter HTT primary reflector which allows for optimum optical coupling to the \pb{} receiver, reducing aberrations and systematics. 
One polarizing grid is placed at the output of the PB-FTS, and modulated via continuous rotation. This modulation allows for decomposition of the signal into different harmonics that can be used to probe potentially pernicious sources of systematic error in a polarization-sensitive instrument.
The high throughput and continuous output polarizer modulation features are unique compared to other FTS calibrators used in the CMB field.
In-field characterization of the \pb{} receiver was accomplished using the PB-FTS in April 2014. We discuss the design, construction, and operation of the PB-FTS and present the spectral characterization of the \pb{} receiver. We introduce future applications for the PB-FTS in the next-generation CMB experiment, the Simons Array.
\end{abstract}

\maketitle

\section{Introduction}
\label{SEC_intro}

In recent years, galactic foregrounds have been recognized as an important source of confusion for cosmic microwave background (CMB) polarization experiments\cite{BKP2015,Kamion2016}.
These astrophysical sources have been, and may continue to be, misinterpreted as the imprint of cosmic B-mode polarization, the so-called ``smoking gun'' of cosmological inflation. It is thus critical to obtain both precise and accurate characterization of the frequency response of CMB experiments.
It is a standard practice for experiments to trade-off between compact spectral characterization sources that are narrow bandwidth, and thus of limited utility for simultaneous foreground and CMB detection, or physically larger spectrometers that can more accurately characterize the instrument performance over the large wavelength regimes needed to mitigate galactic foregrounds.

Of course, instrument and detector characterization is a crucial step in any astrophysical experiment.
Calibrating the experiment ensures the correct frequency bands are observed, constrains any coupling between the detectors and atmospheric emission lines, and quantifies the amplitudes of astrophysical continuum emission. In addition to these benefits, spectral characterization of the instrument and detectors is of particular importance when conducting polarization-sensitive observations of the CMB in that it enables the minimization of temperature-to-polarization leakage --- a systematic error that arises when a polarization signal is extracted from two orthogonally oriented polarization-sensitive detectors. With proposed future ground-based CMB experiments (e.g. Simons Observatory, CMB-S4) aiming for lower noise measurements of inflationary gravitational waves, it is necessary to achieve more precise and accurate spectral characterization of the temperature-to-polarization leakage, atmospheric emission, and foreground (dust and synchrotron) emission systematics.

\pb{} is a CMB polarization experiment that began observations in early 2012.
The \pb{} receiver is installed on the Huan Tran Telescope (HTT) located in the Atacama Desert in Chile at an altitude of 5,200 m. The \pb{} receiver contains 1,274 polarization-sensitive transition edge sensor (TES) bolometric detectors operating within the 150 GHz atmospheric window\cite{KArnold2012,Kermish2012,Barron2014}. 
A direct measurement of the CMB B-mode power spectrum, $C_{\ell}^{BB}$, between $500<\ell<2100$ has been made from the \pb{} first and second season observations\cite{PB2014BB, PB2017}. 

Fourier Transform Spectrometers (FTS) have been widely used in the field of astronomy and for CMB experiments. A Martin-Puplett FTS\cite{Martin1982, Martin1970} can be used to directly observe the CMB signal as a differential instrument that allows for intrinsic common-mode subtraction\cite{Alessandro2015} such as in the COBE-FIRAS\cite{COBE1998} instrument and in the proposed PIXIE\cite{PIXIE2014} instrument. In a related application, a Martin-Puplett FTS has been used to assess the common-mode subtraction performance of the balloon-borne telescope OLIMPO\cite{Alessandro2015}.
Typically for ground-based CMB experiments, a Martin-Puplett FTS is used not as the observing instrument itself, but as a spectral calibrator for characterizing the spectral response of the instrument such as in the Atacama Cosmology Telescope (ACT)\cite{Datta2016}, SPT-3G\cite{Pan2018}, and the Keck Array and BICEP3\cite{BK2014FTS} instruments. 
Similarly, a FTS is used as a spectral calibrator for the instrument in \pb{}.

Spectral response measurements of a subset of the detectors in the \pb{} receiver were performed before deployment\cite{KArnold2012}. However, because the receiver is optically designed to couple directly to the 2.5 m-aperture HTT reflector system, the coupling between the receiver and the FTS used to perform these initial lab measurements was insufficient for obtaining data of the quality needed for CMB analyses for a large number of pixels within the focal plane. This motivated the fabrication of a custom-built high-throughput FTS.

Improved characterization of the \pb{} instrument and its detectors was performed in April 2014 using a FTS that was specially designed for use with the \pb{} receiver while on the HTT. This \pb{}-FTS (PB-FTS) is a Martin-Puplett interferometer which mounts directly to the HTT between its primary and secondary mirrors. 
The PB-FTS is more optimally coupled to the \pb{} receiver compared to the laboratory FTS and can thus produce spectra with higher signal-to-noise ratio (SNR) and less systematics more efficiently.
Compared to other FTS calibrators used within the CMB field, the PB-FTS is unique in that it has higher throughput and employs a continuously rotated output polarizer. It has also been designed to optimally couple to the \pb{} receiver and HTT through the implementation of a custom made output parabolic coupling mirror.

In this paper we describe the theory, design, and construction of the PB-FTS and present results taken in the field from \pb{}. The basic FTS theory and continuous modulation theory are explained in Section \ref{SEC_background}. The optical design and construction of the PB-FTS components are explained in Section \ref{SEC_instrumentation}. The in-field data and analysis results from \pb{} are shown in Section \ref{SEC_dataanalysis}. 
Discussion and future applications for this PB-FTS are described in Section \ref{SEC_future}. 

\section{Fourier Transform Spectrometer}
\label{SEC_background}

\subsection{Martin-Puplett Interferometer}\label{SUBSEC_MPI}

\begin{figure}[ht]
\centering
\includegraphics[width=.5\textwidth]{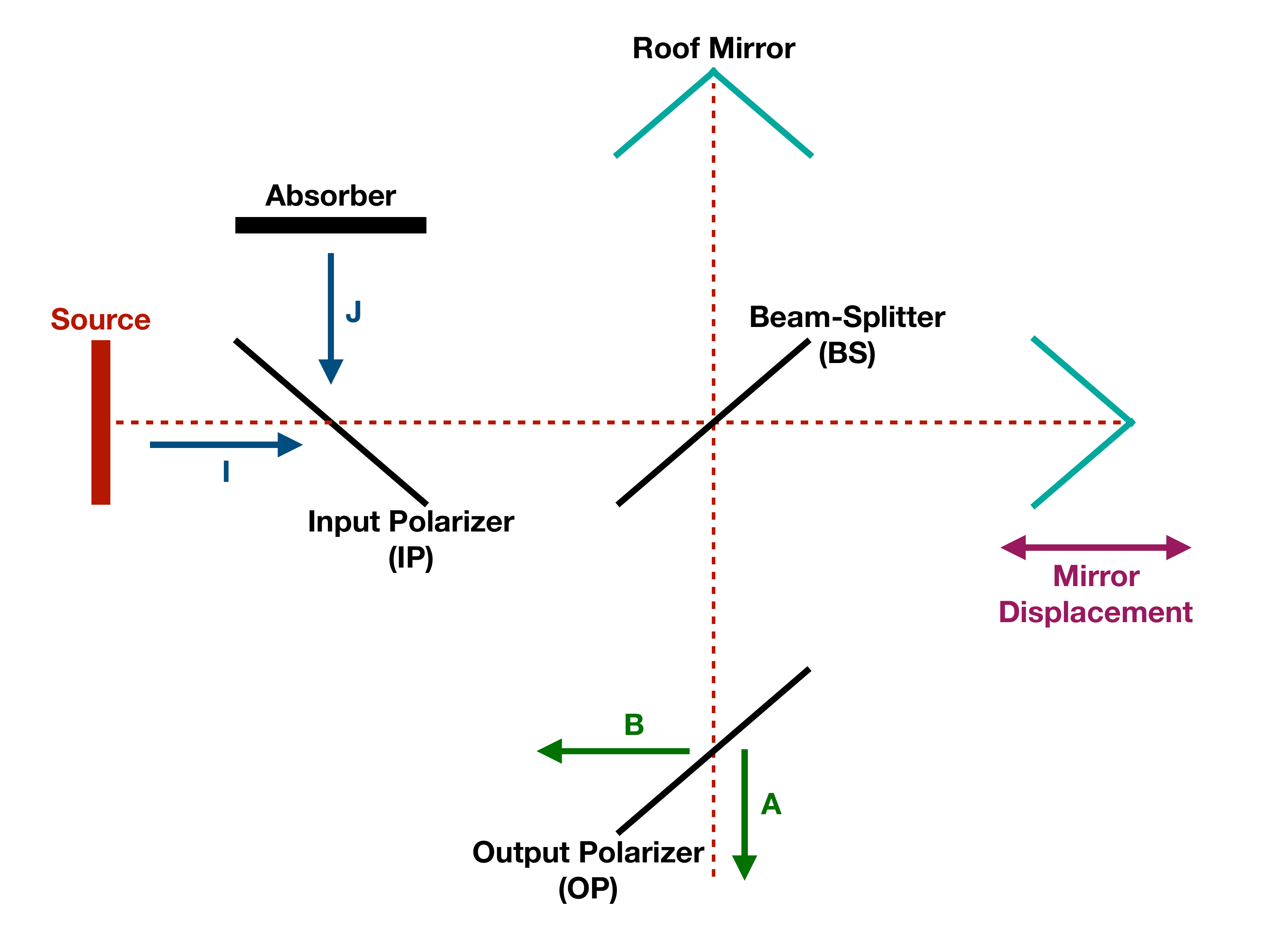}\hfill
\caption[Schematic of Martin-Puplett interferometer]{\label{FIG_FTSSchematic} A schematic drawing of the Martin-Puplett interferometer. $I$ is the input signal from the thermal source placed at one input port, and $J$ is a blackbody absorber placed at the second input port. $A$ and $B$ are the two output ports of the interferometer.}
\end{figure}

The PB-FTS is a Martin-Puplett interferometer \cite{Martin1982, Martin1970}. A Martin-Puplett interferometer is based on a Michelson interferometer but differs in three ways: the beam-splitter is a wire grid polarizer, the mirrors which reflect the two interfering beams are roof mirrors that rotate the polarization of reflected light by 90 degrees, and the output of the interferometer is polarized. 
A schematic illustration of a generic Martin-Puplett interferometer with a source at one input port is shown in Figure \ref{FIG_FTSSchematic}. 
A phase difference is introduced into the interferogram signal by controlling a movable roof mirror in one of the two beams (``arms'').

A Martin-Puplett interferometer has two advantages over a standard Michelson interferometer. The first is that the Martin-Puplett interferometer has efficient beam-splitting with theoretically no frequency dependence in transmission or reflection due to the wire grid polarizers. The second advantage is that the Martin-Puplett interferometer is a four-port device with two input and two output ports. The output ports provide complementary signals in which simple subtraction or continuous modulation by rotating the output polarizer allows for suppression of spurious noise in the signal and unpolarized source. The continuous modulation set-up has some unique characteristics in the output signal as explained in Section \ref{SUBSEC_mod}.

The PB-FTS uses a thermal source that is polarized by the input wire grid polarizer placed in its diverging beam. An input parabolic mirror collimates the radiation from the source and directs it toward the beam-splitter. An identical output parabolic mirror focuses the output beam, and the output rotating wire grid polarizer is placed near its focus. The specific design of the PB-FTS is described in Section \ref{SUBSEC_design}.

\subsection{Modulation Theory}\label{SUBSEC_mod}

For \pb{}, rather than using a standard modulation technique such as chopping the source, the output polarizer is continuously rotated in order to modulate the signal. With this technique, the measured signal can be decomposed into various harmonics of the rotation frequency. 
The basic concept of modulation in a Martin-Puplett interferometer utilizing the input and output polarizers has been studied by Martin\cite{Martin1982} and Da Costa\cite{DaCosta1990}.
Using the Jones formalism and the methodology from Martin\cite{Martin1982} as a basis, the various dependencies on the polarizer rotation and detector polarization angle can be further derived for the specific case of coupling to a polarization-sensitive instrument.

In a Martin-Puplett interferometer, only relative angles between the polarizers and roof mirrors are important. 
An overall rotation of the whole optical system is unimportant. Therefore the Jones formalism methodology presented by Martin will apply for all generic Martin-Puplett interferometers. In the case of a polarization-sensitive instrument like the \pb{} receiver, an additional polarizer matrix must be included in the calculation to represent the polarization-sensitive axis of the polarized detector. 

The electric field amplitude through the interferometer output port $A$ coupled to a polarization-sensitive detector is given by the following matrix calculation:
\begin{eqnarray}
E_{\rm det} & = & R_{\rm det}(\alpha-\theta(t)) A(\theta(t)) \nonumber \\
& = & R_{\rm det}(\alpha-\theta(t)) T_{\rm OP} R_{\rm OP}(\theta(t)) D E_{\rm IP}.
\end{eqnarray}
The amplitude from the interferometer output port $A$ is given by $A = T_{\rm OP} R_{\rm OP} D E_{\rm IP}$ and the amplitude at the detector is $E_{\rm det}$.
$E_{\rm IP}$ is the amplitude from the input polarizer that combines the amplitudes $I$ and $J$ from the interferometer input ports.
$D$ is the matrix introducing the phase shift between the two arms, $T_{\rm OP}R_{\rm OP}$ represents the transmission through the output polarizer, and $R_{\rm det}$ represents the polarized detector. $\theta(t)$ and $\alpha$ represent the output polarizer angle and detector polarization-sensitive angle, respectively. $\theta(t)$ has a time dependence and represents the continuous modulation. Here it is assumed that the beam-splitter is aligned 45 degrees relative to the input polarizer. Therefore the matrix $D$ implicitly represents the process of beam-splitting, phase shift, and beam-combining. Calculating the signal power at the detector, one obtains 
\begin{equation}
E_{\mathrm{det},T} E_{\mathrm{det},T}^{\ast} = E_{\rm DC}^{2} + E_{2f}^{2} + E_{4f}^{2} 
\end{equation}
\begin{eqnarray}
E_{\rm DC}^{2}  & = & a_{1} + \frac{1}{2} a_{2} \cos \left( 2 \alpha \right) +  \frac{1}{2} a_{3} \sin \left( 2 \alpha \right) \nonumber \\
& = & \frac{1}{2} a_{2} (\Delta) \cos \left( 2 \alpha \right) + C_{\rm DC} \label{eqn_DCterm} \\
E_{2f}^{2}  & = & a_{1} \cos \left( 2 \theta(t) - 2 \alpha \right) + a_{2} \cos \left( 2 \theta(t) \right) + a_{3} \sin \left( 2 \theta(t) \right) \nonumber \\
& = & a_{2} (\Delta) \cos \left( 2 \theta(t) \right) + C_{2f} \label{eqn_2fterm} \\
E_{4f}^{2}  & = & \frac{1}{2} a_{2} \cos \left( 4 \theta(t) - 2 \alpha \right) + \frac{1}{2} a_{3} \sin \left( 4 \theta(t) - 2 \alpha \right) \nonumber \\
& = & \frac{1}{2} a_{2} (\Delta) \cos \left( 4 \theta(t) - 2 \alpha \right) + C_{4f} \label{eqn_4fterm}
\end{eqnarray}
where the signal consists of DC, $2f$ modulated, and $4f$ modulated terms. The subscript $T$ represents the transmitted component through the detector's polarization-sensitive axis. The phase difference between the two interfering beams is given by $\Delta = kx$ where $k$ is the wavenumber and $x$ is the path length difference.
Any terms labeled with $C$ are constant values after demodulation. The $a_{i}$ terms are
\begin{eqnarray}
a_{1} & = & \frac{1}{8} \left( d_{f}^{2} + d_{m}^{2} \right) \left( I_{s}^{2} + J_{p}^{2} \right) \\
a_{2} & = & \frac{1}{4} \left( I_{s}^{2} - J_{p}^{2} \right) \cos \left( \Delta \right) \\
a_{3} & = & \frac{1}{8} \left( d_{f}^{2} - d_{m}^{2} \right) \left( I_{s}^{2} + J_{p}^{2} \right) .
\end{eqnarray}
The interferogram terms of interest are those that only depend on $a_{2}$ which contains the $\Delta$ dependence. $d_{f}$ and $d_{m}$ are the complex propagation coefficients for the fixed mirror and moving mirror arms. They represent the loss in amplitude of the signal as they propagate through the two arms. $I$ and $J$ are the signals from the input ports and the subscripts $s$ and $p$ represent the orthogonal amplitude components parallel and perpendicular to the input polarizer wires, respectively. A detailed derivation of these results is shown in the appendix.

The DC term contains the interferogram signal dependent on the detector polarization angle.
The $2f$ term contains the interferogram signal independent of the detector polarization angle. The $4f$ term contains an interferogram signal that depends on the relative angle between the output polarizer angle and detector polarization angle. Hence, theoretically even without knowing $\alpha$ or the relative angle between $\theta$ and $\alpha$, the spectrum can be measured using the $2f$ signal. 
The $2f$ interferogram signal is a factor of two stronger in magnitude compared to the DC and $4f$ interferogram signals. 
The DC signal is not modulated and typically will be difficult to measure due to the $1/f$ noise of the instrument and environment. Therefore when performing the spectral analysis of the \pb{} instrument, the $2f$ modulated interferogram signal is primarily used. 
This separation of signals into different harmonics (up to a $4f$ component) with each harmonic having different parameter dependencies only arises when applied to a polarization-sensitive instrument. 
As will be mentioned in Section \ref{SEC_future}, this $4f$ modulated signal has possibility for further interesting analyses for probing the variation of polarization-axis sensitivity as a function of frequency.

The analysis presented, which applies to a Martin-Puplett interferometer coupled directly to a polarization-sensitive detector, also applies to the full \pb{} receiver, which contains additional optical elements between the PB-FTS and the detector. These include lenses, filters, and a half-wave plate (HWP). Even though each of these elements has its own polarization properties, they do not actively modulate the incoming polarization signal during the FTS measurement. The HWP was not stepped from mid-first season and throughout second season of observations\cite{PB2017}, and likewise was not stepped during the FTS measurements. Therefore the PB-FTS measurements were taken under the same conditions as during CMB observations and characterize the spectral response of the \pb{} instrument under typical conditions. Because the output polarizer is the only source of active modulation, theoretically the PB-FTS signal observed by the \pb{} instrument should show the same dependencies on $\theta(t)$, $\alpha$, and $\Delta$ as in the results above.

\section{Instrument Design and Hardware}
\label{SEC_instrumentation}

\subsection{Instrument Design}\label{SUBSEC_design}

The PB-FTS was specifically designed to measure the spectral response of the installed \pb{} receiver instrument and its detectors in the field and thus had to operate while installed on the HTT with the \pb{} receiver in place. This requirement put several constraints on its design. In particular, the PB-FTS had to be relatively lightweight, able to withstand environmental conditions in the Atacama Desert in Chile, and capable of optically coupling to the \pb{} receiver efficiently.

Additionally, high-efficiency optical coupling between the FTS and the receiver was desired in order to reduce systematic errors.
If the FTS signal does not effectively fill the beam of the detectors in the receiver, the measured spectra are more prone to potential systematic errors such as those arising from a miscorrection of the frequency dependence in the detected power\cite{BK2014FTS}. If a detector is only partially beam filled by the FTS signal, this signifies that the detector is simultaneously seeing light from a different source, for example, such as stray light entering into the receiver due to reflections off the FTS enclosure and mount structures. In the field, these stray light systematics are typically difficult to control. Because the telescope and receiver optics are designed such that the detectors are effectively beam filled by the sky signal during regular observations, any FTS measurement that cannot effectively beam fill in a similar fashion can suffer from increased systematics that are not present during regular observations. It also has been empirically observed in laboratory measurements that insufficient beam filling of a detector can cause large systematic fluctuations in the in-band spectrum. 

High-efficiency optical coupling is also desired in order to increase measurement efficiency.
The larger the number of pixels that the FTS can simultaneously measure, the smaller the number of separate FTS measurement runs required. Thus the entire focal plane array can be characterized in a shorter amount of time.

Therefore the PB-FTS is desired to be capable of effectively fully illuminating multiple pixels within the receiver simultaneously.  This is achieved by designing an optical system with a throughput of 15.1 steradian cm${}^{2}$ and utilizing an output parabolic mirror designed specifically for use with the HTT. 

\begin{figure*}[htp]
\centering
\raisebox{-0.5\height}{\includegraphics[width=.5\textwidth]{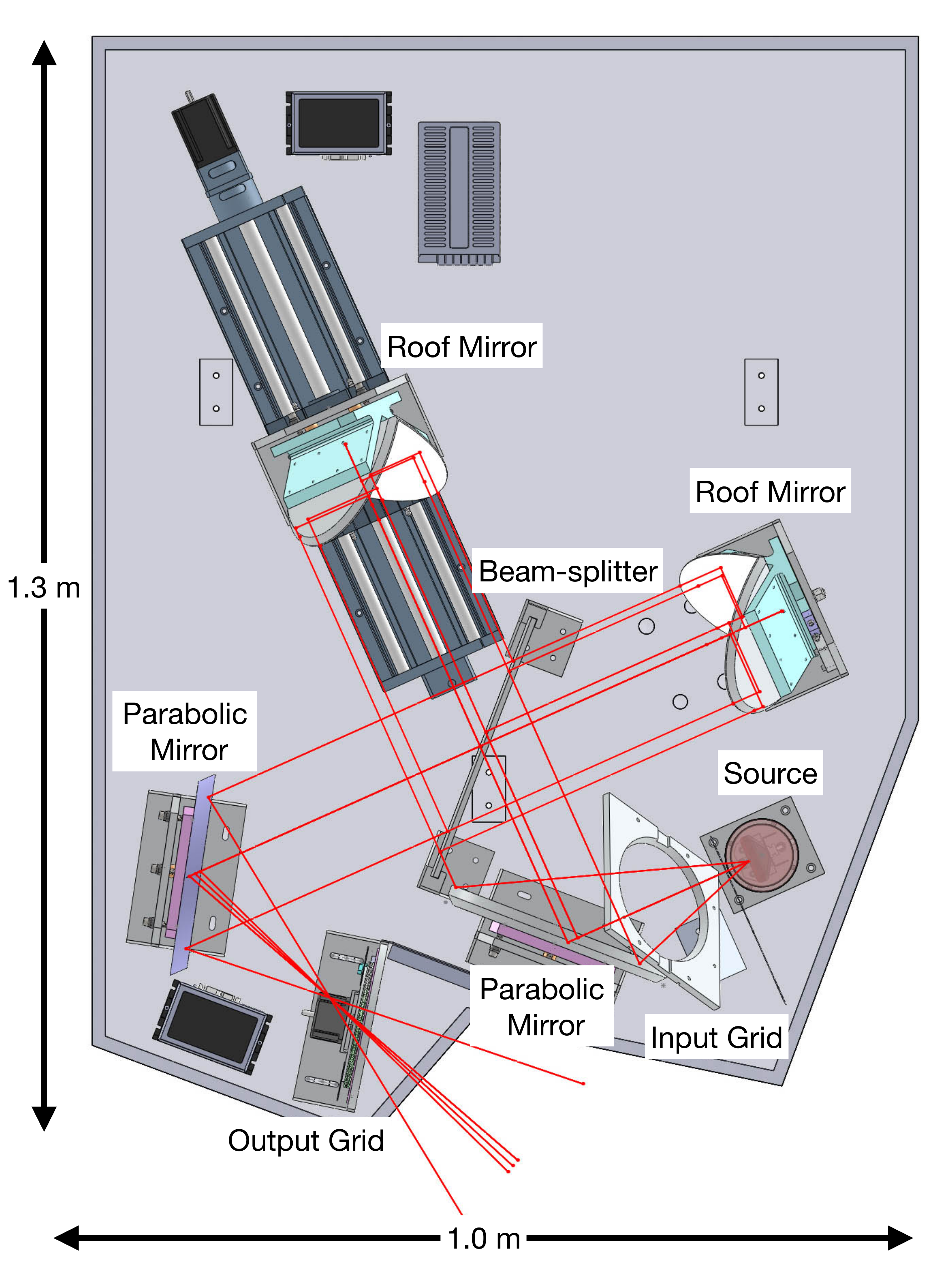}}\hfill
\raisebox{-0.5\height}{\includegraphics[width=.5\textwidth]{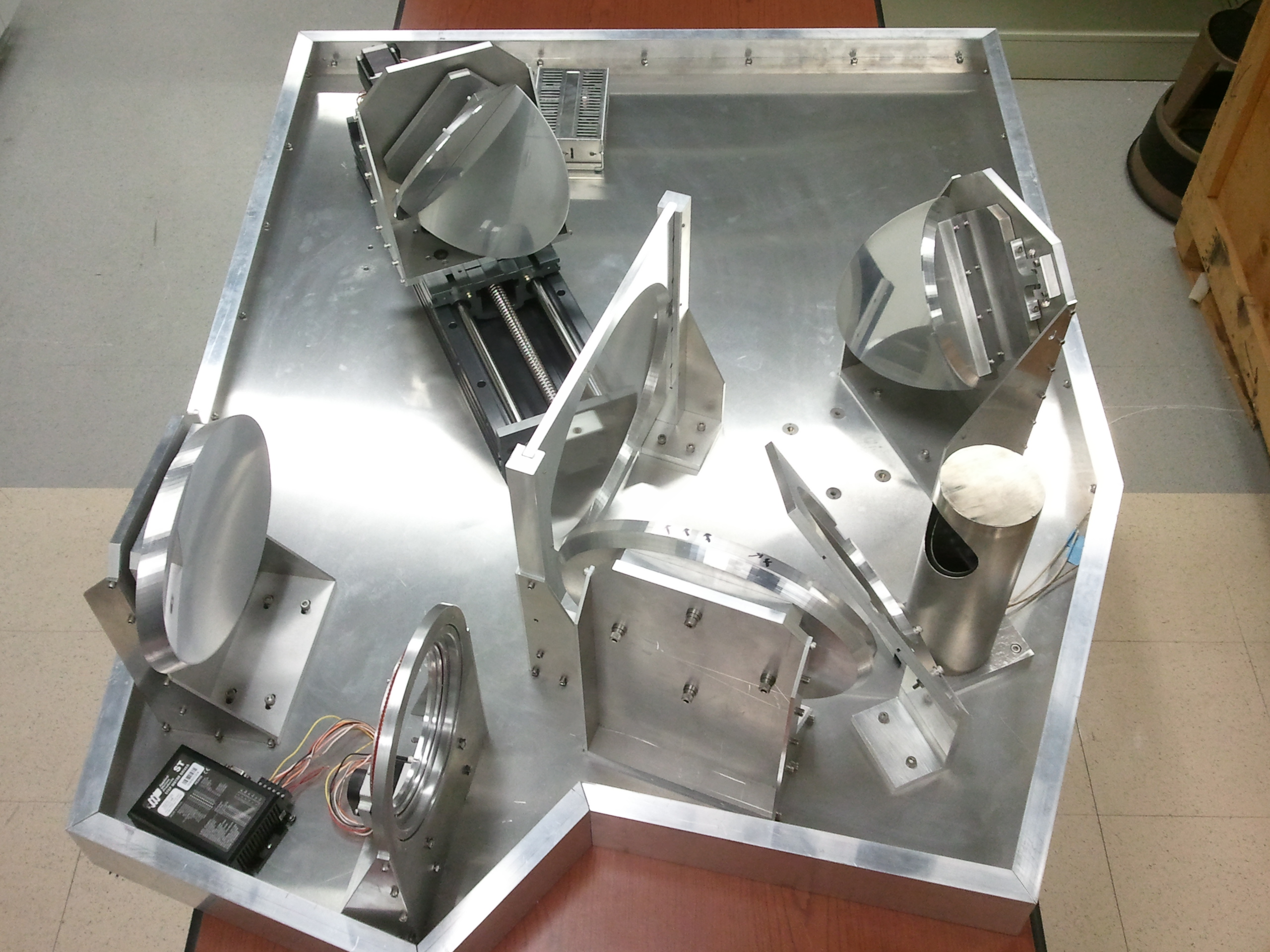}}\hfill
\caption[Mechanical design of PB-FTS]{\label{FIG_FTSChile} The mechanical drawing of the PB-FTS is shown on the left, and an image of the completed PB-FTS is shown on the right. The enclosure and partition walls are not shown in the image to highlight the PB-FTS optical components. In the deployed PB-FTS, all walls and optical bench surfaces were covered with blackbody absorber.}
\end{figure*}

The overall PB-FTS design is shown in Figure \ref{FIG_FTSChile}. It consists of a T-SHTS/4 ceramic heater source made by Elstein\cite{Elstein}, two parabolic mirrors to collimate and then focus the radiation from the source, one fixed roof mirror and one movable roof mirror mounted to a linear translation stage, and three wire grids used as input, output, and beam-splitting polarizers. The heater source radiating area is 60 mm $\times$ 60 mm and can be heated up to 900 $^{\circ}$C. The mirrors are made with MIC-6 aluminum and the reflecting surfaces are machined to a surface flatness RMS of $<25$ microns in order to reduce any power loss resulting form surface irregularity to $<2.5\%$ at 150 GHz, as derived from the Ruze criterion\cite{Ruze1966}.  Further details about the design and fabrication of the parabolic mirrors and wire grid polarizers are discussed in the Sections \ref{SUBSEC_mirrors} and \ref{SUBSEC_grids}.  All components are mounted to an optical bench constructed of aluminum honeycomb in order to minimize weight while maintaining strength. Apart from an aperture at the output, the entire optical system is surrounded by an aluminum enclosure whose inner walls are lined with blackbody absorber in order to minimize the amount of stray light that can enter the optical path, as well as terminate reflected light from the polarizers.

All PB-FTS optical components were designed and machined at the University of California, San Diego, and all polarizing wire grids were fabricated in-house using our custom wire gird winder.

In order to achieve the throughput as stated above, all optical components within the collimated main beam of the FTS have a 20 cm-diameter clear aperture. The source size was also optimized to fully utilize this 20 cm clear aperture when coupled using the input parabolic mirror. The PB-FTS's large aperture size and throughput make it unique compared to other field-use FTSs, many of which are designed to be physically compact and whose optical designs are thus more constrained\cite{Datta2016,Pan2018,BK2014FTS}.

The optical and physical specifications of the PB-FTS are listed in Table \ref{TABLE_specs}. The PB-FTS is capable of operation in two modes that allow for different spectral resolutions.  
The length of the fixed roof mirror arm can be adjusted to move the location of the maximum constructive interference.
During operation in mode 1, the fixed roof mirror is positioned at a distance from the beam splitter equal to the distance between the movable roof mirror and the beam splitter when the movable mirror is at the midpoint of the linear stage. In this mode a double-sided interferogram can be measured as the movable mirror is translated across the length of the linear stage and allows for a 1 GHz spectral resolution measurement. During operation in mode 2, the fixed mirror is positioned closer to the beam splitter, at a distance equal to the minimum distance between the beam splitter and the movable mirror. In this mode the resulting spectrum will have 0.5 GHz resolution, a factor of two better compared to mode 1, but it is only possible to measure a single-sided interferogram. Because single-sided interferograms are less sensitive to potential systematic effects such as misalignment of optics, unwanted reflections, and partial detector illumination, mode 1 is the default mode for use with \pb{}.

\begin{table}
\begin{center}
\caption[PB-FTS specifications]{\label{TABLE_specs} The designed specifications of the PB-FTS are listed. The PB-FTS can be used in two different modes that change the interferogram type and frequency resolution in the calculated spectrum. Any specifications that do not distinguish between the two modes are common.}
\begin{tabular}{c|c}
\hline
\hline
Specifications & Mode 1 (Mode 2) \\
\hline
Interferogram & Double (Single) \\
Frequency Resolution & 1 GHz (0.5 GHz) \\
Maximum Frequency & 500 GHz \\
Throughput & 15.1 steradian cm$^{2}$ \\
Output f-number & $f=1$ \\
Dimensions & 1.3 m $\times$ 1.0 m \\
Weight & 70 kg \\
\end{tabular}
\end{center}
\end{table}

\begin{figure*}[htp]
\centering
\includegraphics[width=.65\textwidth]{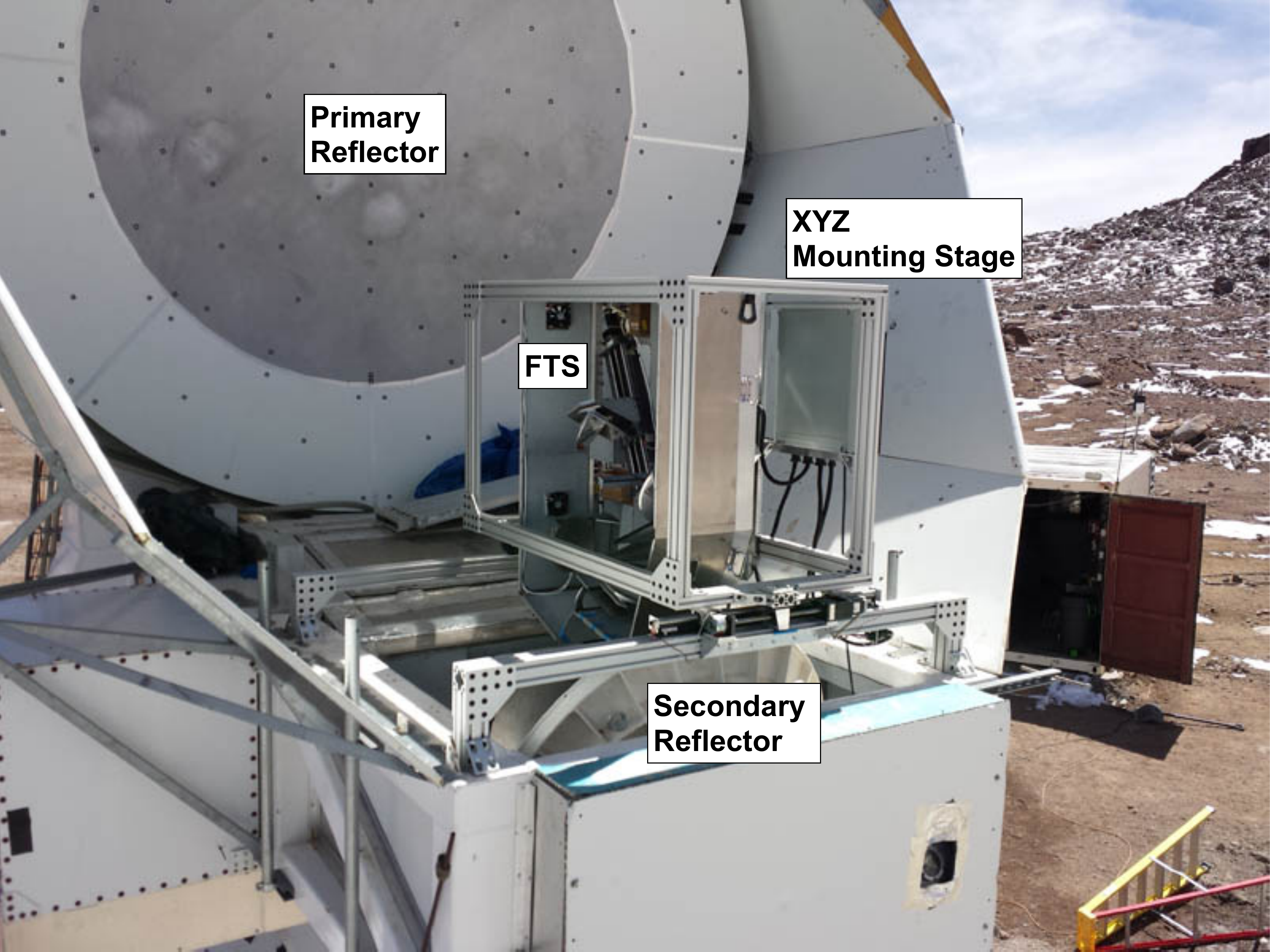}\hfill
\caption[XYZ mounting stage]{\label{FIG_XYZStage} An image of the XYZ stage mounted on the HTT boom with the PB-FTS installed in Chile is shown.}
\end{figure*}

The PB-FTS mounts on top of the HTT lower boom structure at the prime focus located between the primary and secondary reflectors using an XYZ mounting stage\cite{8020Inc}, shown in Figure \ref{FIG_XYZStage}. The prime focus plane is a plane perpendicular to the central optical axis in which the \pb{} focal plane geometric rays come to an approximate focus and field stop. The mounting stage contains three linear translation stages that allow the PB-FTS to be positioned at different locations around the prime focus plane, thereby illuminating different parts of the \pb{} detector focal plane. With this mounting stage, the PB-FTS can move $\pm 10$ cm in both X and Y directions within the prime focus plane as well as $\pm 5$ cm perpendicular to this plane in order to efficiently couple to all the pixels in the focal plane. All three translation stage can be remotely operated to allow for automatic scanning and data taking.

\subsection{Parabolic Mirrors and Optical Coupling}\label{SUBSEC_mirrors}

The PB-FTS employs two parabolic mirrors: one to collimate the beam from the source and another to focus the FTS output signal. The two parabolic mirrors have identical conical shapes in order to preserve the f-number in the input and output ports. This shape was designed to act like a small version of the HTT primary reflector, and specifically the output parabolic mirror is used for optically coupling the PB-FTS to the \pb{} receiver.

The HTT reflectors satisfy the Mizuguchi-Dragone condition \cite{Dragone1978,mizuguchi1976}, thereby minimizing astigmatic aberration as well as limiting cross-polarization at and near the central optical axis of the system. The PB-FTS parabolic mirror is a 65 degree off-axis paraboloid with $f=1$, consistent with the 2.5 m HTT primary reflector shape\cite{Kermish2012}, but much smaller with only a 20 cm diameter clear aperture. Thus the PB-FTS parabolic mirror is equivalent to the primary reflector shrunk down proportionally in all dimensions. 

\begin{figure*}[htp]
\centering
\raisebox{-0.5\height}{\includegraphics[width=.45\textwidth]{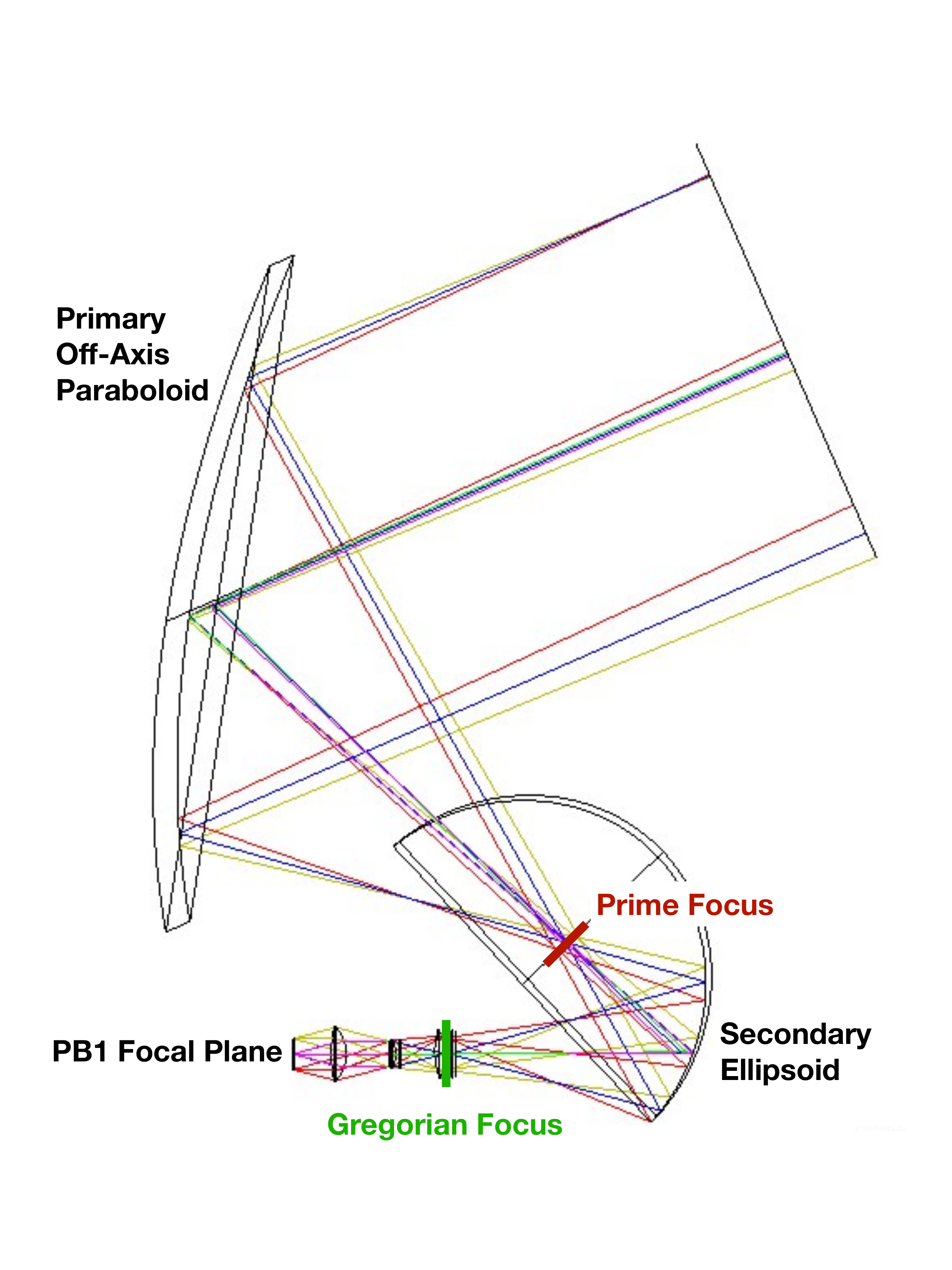}}\hfill
\raisebox{-0.5\height}{\includegraphics[width=.55\textwidth]{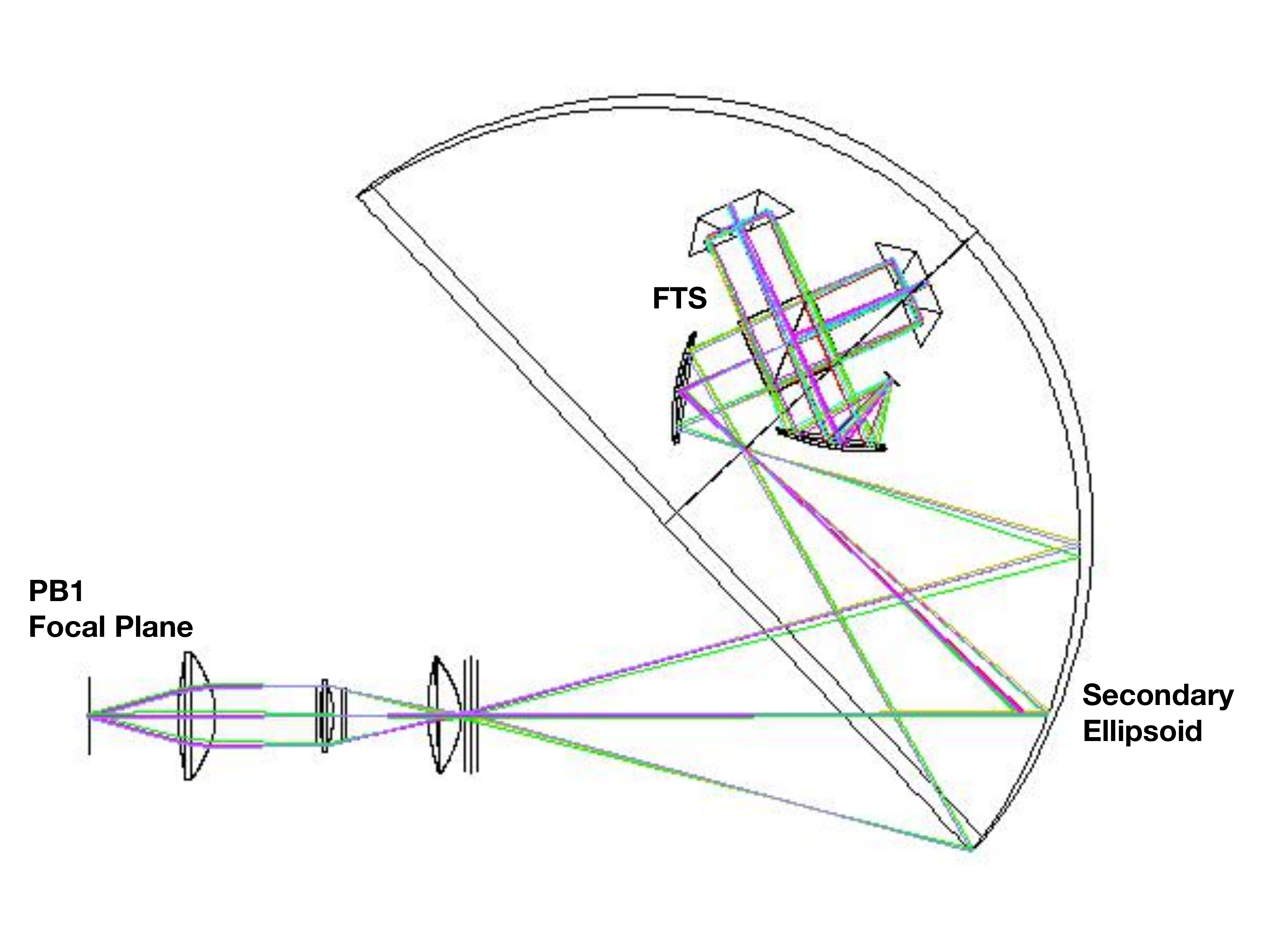}}\hfill
\caption[PB-FTS optical coupling to \pb{}]{\label{FIG_MDParabola} Geometric ray traces of the full \pb{} optics (left) and the PB-FTS coupled to the secondary mirror and receiver (right). In both cases the Mizuguchi-Dragone condition is satisfied. The PB-FTS as shown here couples to 19 pixels at once.}
\end{figure*}

When the PB-FTS is properly aligned and mounted on the HTT, the output PB-FTS parabolic mirror and HTT secondary reflector also satisfy the Mizuguchi-Dragone condition. Hence along the central optical axis of the PB-FTS and \pb{} system, the astigmatism is canceled. 
This orientation is shown in Figure \ref{FIG_MDParabola}. The PB-FTS is placed between the primary and secondary reflectors around the prime focus plane. Due to the smaller size of the PB-FTS parabolic mirror, the PB-FTS is placed closer to the secondary and couples to a smaller portion of the detector focal plane detectors at once. Coupling to different parts of the focal plane is achieved using the XYZ mounting stage as discussed earlier. 

Theoretically, increasing the optical diameter of the parabolic mirror and throughput of the FTS allows for simultaneous optical coupling to a larger number of focal plane detectors. In the limit that the parabolic mirror nears in size to that of the HTT primary reflector, the entire focal plane can be coupled at once without any positional adjustment of the FTS. In order for the rays from multiple focal plane detectors to reach the FTS source unvignetted, the FTS throughput itself must also be increased accordingly which is equivalent to making the collimated main beam aperture size within the FTS larger. The 20 cm diameter clear aperture of the PB-FTS (parabolic mirrors and optical components in the FTS collimated beam) was chosen such that the PB-FTS is able to simultaneously couple to and effectively beam-fill a reasonable number of $\sim$19 detectors at once with the installed thermal source while keeping the entire size of the PB-FTS at an implementable and manageable level. This throughput in combination with the parabolic mirror shape 
allows for reasonably efficient optical coupling of the PB-FTS and \pb{} receiver.

From Zemax\cite{ZEMAX} optical ray tracing simulations it was found that, near the central axis of the system, the optical coupling between the receiver and PB-FTS is efficient. When the PB-FTS is aimed at the center pixel, greater than 95\% of the equally spaced geometric rays that pass through the \pb{} Lyot stop are unvignetted and reach the PB-FTS source (reverse time frame). The root mean square (RMS) wavefront error when coupled is simulated to be less than 1\% of one wavelength. 
It is expected that the optical coupling efficiency will decrease as a function of distance from the central optical axis when aiming the PB-FTS at different parts of focal plane.
From simulations with the PB-FTS targeting the outermost edge pixels, greater than 87\% of the rays are unvignetted and the RMS wavefront error is less than 25\% of one wavelength at 150 GHz. Despite this decreased performance for edge pixels, this design should theoretically perform better on average and maintain high efficiency optical coupling across the focal plane compared to other standard parabolic mirror shapes that do not satisfy the Mizuguchi-Dragone condition. 

From these simulations, it was also found that tilting of the PB-FTS was not necessary in order to efficiently couple across the focal plane. This is because the \pb{} optical rays only come to an approximate focus near the prime focus plane. Only the central pixel comes to a sharp focus and the spot sizes of the focal plane pixels increase as a function of distance from the central pixel. While there theoretically exists an optimal coupling location for the central pixel, there is no theoretical optimum for the rest of the focal plane pixels. Because equally efficient coupling across the focal plane pixels could be obtained in simulations with only the XYZ motions, it was decided to not implement a tilting mechanism in order to greatly simplify the mounting stage. 

\subsection{Polarizing Grids}\label{SUBSEC_grids}

The FTS contains three types of polarizing grids: an input grid, a beam-splitter grid, and an output grid. The necessity to keep the PB-FTS reasonable in size required three different grid sizes and shapes in order to utilize space efficiently. All the grids are strung with 25 micron-diameter tungsten wire at 100 micron spacing. Theoretically for tungsten wires (electrical resistivity $5.5 \times 10^{-8} \ohm \cdot$ m) of finite diameter, wire grids with this specification can achieve polarization efficiencies of 99.95\% in transmission and 99.76\% in reflection for normal-incidence light at 150 GHz \cite{Manabe2005,Larsen1962}. Even at 170 GHz, near the top of the \pb{} band, the wire grids can theoretically achieve polarization efficiencies of 99.94\% in transmission and 99.72\% in reflection.

\subsubsection{Grid Winder}\label{SUBSEC_gridwinder}

The collimated portion of the PB-FTS instrument is designed to have a 20 cm diameter clear aperture in order to provide the desired throughput. 
Polarizing grids of this larger size with high polarization efficiency across the \pb{} frequency range are difficult to purchase at reasonable cost. Commercially available polarizing grids are typically either too small in size or expensive because they are developed for use at higher frequencies. 
Hence, a grid winder was developed for \pb{} and used to fabricate all polarizers employed by the PB-FTS.

\begin{figure}[htp]
\centering
\raisebox{-0.5\height}{\includegraphics[width=.5\textwidth]{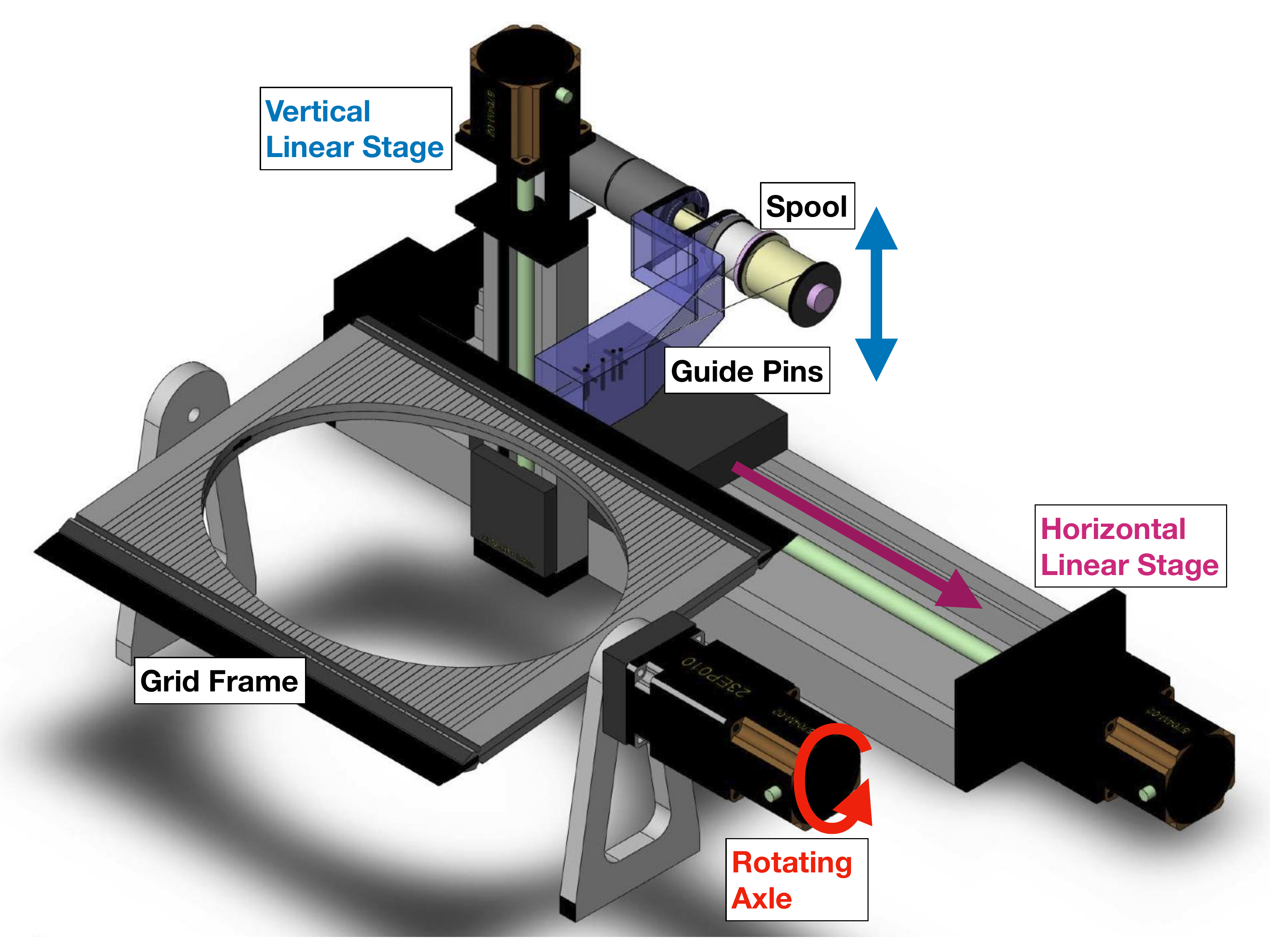}}\hfill
\caption[Custom polarizing wire grid winder design]{\label{FIG_GridWinder} The mechanical design of the custom polarizing wire grid winder is presented. The motions of the vertical linear stage (blue arrow), horizontal linear stage (magenta arrow), and rotating axle (red arrow) are shown. For each full revolution of the axle, the vertical linear stage moves the spool and guide pins mechanism back-and-forth (between the highest and lowest vertical positions) twice. This synchronizes the height of the spool and guide pins mechanism with the winding edge of the grid frame at all times. This grid winder was built at the University of California, San Diego and has been functioning for several years. It was used to fabricate all the grids in the PB-FTS.}
\end{figure}

The grid winder is shown in Figure \ref{FIG_GridWinder}. It is composed of two linear stages (one vertical and one horizontal) and one rotating axle. The frame on which the wire grid is to be wound is mounted on the rotating axle and rotated slowly at a constant rate. A tungsten wire spool is mounted on the vertical linear stage whose motion and vertical height is synchronized with that of the winding edge of the rotating grid frame. As the grid frame rotates, the spool moves vertically such that the wire getting wound on to the grid frame is always held parallel to the ground. This keeps the angle at which the wire comes off the spool fixed at all times, and minimizes any variations in the tension in the wire throughout the winding process. The vertical linear stage is mounted on top of the horizontal linear stage that slowly moves in the direction perpendicular to the wires, controlling the wire spacing.

A controllable magnetic clutch applies back tension on the spool as the wire is being spun off to minimize wire sag. Tension is set at 45 gram-force. Guide pins are installed to keep the wire between the spool and grid aligned at all times. A small voltage difference is created across two of the pins that are in contact with the wire at all times during the winding. If the wire breaks during the winding process, the device will sense an ``open circuit'' and is programmed to stop. This allows the user to know exactly at what point in time the wire broke so that the winding process can be restarted at that exact location. 

The grid winder allows for precision spacing of the tungsten wire with less than 10 micron error in spacing, and the sag in the wire is less than 20 microns for the largest grid sizes. The device is capable of winding up to 37 cm $\times$ 37 cm rectangular grids. In order to create accurate polarizing grids, the grids are wound very slowly, with each grid typically requiring between 10 and 24 hours to wind depending on the size. One winding process makes two grids of the same design at a time by having two grid frames mounted back-to-back on the axle. Once the winding is complete, epoxy is poured into predesigned grooves in the grid frames to secure the wire onto the grid frames. 

\subsubsection{Beam-splitter and Input Polarizing Grids}
\label{SUBSEC_inputbsgrids}

\begin{figure*}[htp]
\centering
\includegraphics[width=.5\textwidth]{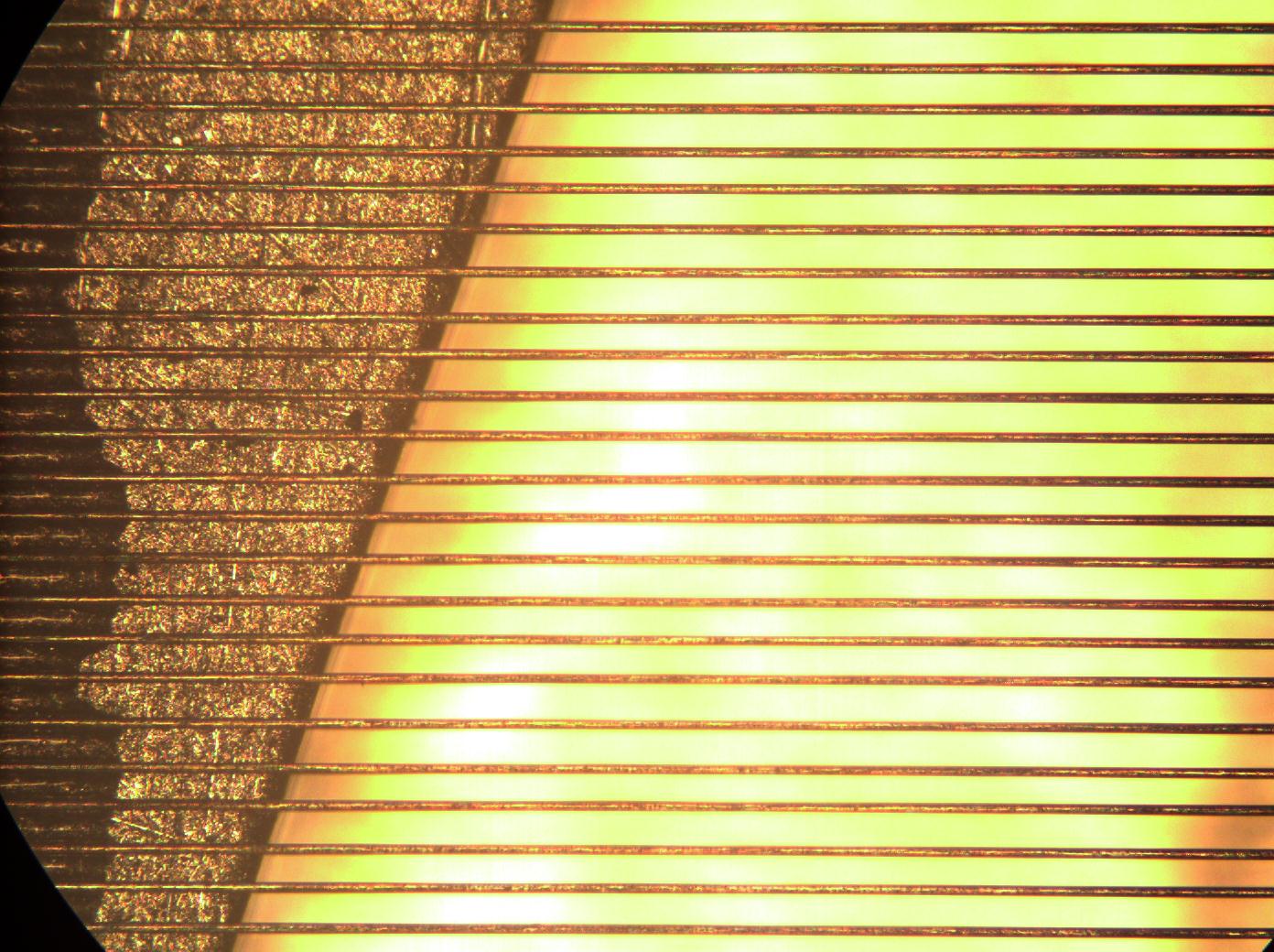}\hfill
\includegraphics[width=.5\textwidth]{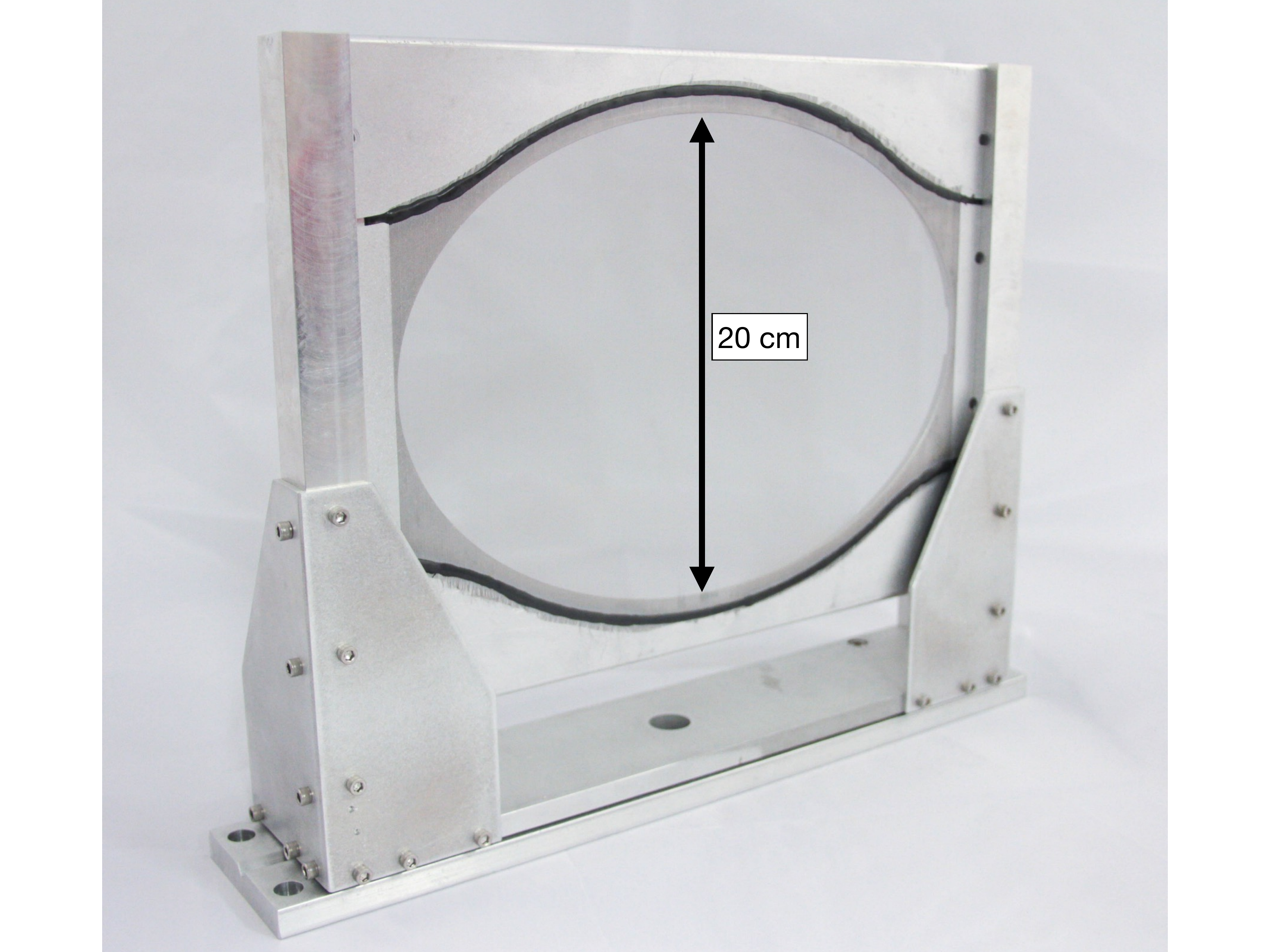}\hfill
\caption[Wire grid polarizer for beam-splitter]{\label{FIG_BeamSplitter} A microscope image of the wire grid wound at 100 micron pitch with 25 micron tungsten wire is shown on the left. The right is the completed beam-splitter wire grid polarizer.}
\end{figure*}

The beam-splitter grid is the largest grid in the system and is shown in Figure \ref{FIG_BeamSplitter}. It is rectangular with a 28.3 cm (major axis) by 20 cm (minor axis) clear elliptical aperture. The aperture is designed such that when placing the grid 45 degrees relative to the collimated beam, the clear aperture in that projection is 20 cm in diameter. The beam-splitter wires are oriented in the normal direction to the optical bench.

The input grid is rectangular and has a 13.3 cm (major axis) by 11.2 cm (minor axis) clear elliptical aperture. It is placed in the diverging beam between the source and first parabolic mirror, and tilted relative to the central ray to ensure that the reflected light is terminated at the enclosure walls layered with blackbody absorber.  
The input grid is oriented such that the wires are oriented 45 degrees relative to the optical bench when viewed in the plane perpendicular to the collimated beam. 

\subsubsection{Rotating Output Polarizing Grid}\label{SUBSEC_outputgrid}

The output grid is circular and has a 12.4 cm-diameter clear aperture. It is placed in the converging beam between the second parabolic mirror and FTS output. Similar to the input grid, the output grid is tilted relative to the central ray. The output grid is mounted to a circular bearing that is continuously rotated by a belt drive and a stepper motor from Applied Motion Products Inc\cite{AppliedMotion} in order to provide the modulation in the output signal. The rotation is measured using an optical interrupter sensor. During data-taking, the output grid was rotated at two revolutions per second.

\section{Data and Results}
\label{SEC_dataanalysis}

The PB-FTS was deployed and installed on the HTT in Chile in April 2014 (after the second season of \pb{} observations). Installation, in-field testing, and full spectral measurements of the \pb{} instrument and detectors across the focal plane were done through the entire month of April 2014. This is near the season in the Atacama region when the precipitable water vapor (PWV) is typically high and makes CMB observations difficult. The PB-FTS successfully took spectral measurements for $\sim$69\% of focal plane detectors in this single deployment. Measurements of all detectors could not be obtained due to limitations from observational time constraints, temporarily inactive detectors, and temporary unexpected telescope data acquisition noise in certain data runs.  

\subsection{Measurements and Data}\label{SUBSEC_data}

The PB-FTS can fully illuminate ${\sim}$19 pixels simultaneously when placed at a single location in the prime focus plane. Thus in order to measure spectral data across the \pb{} focal plane, multiple data runs had to be taken with the PB-FTS at different locations.


The FTS signal is much larger in power than the sky signal for which the detectors are designed. To account for this, the TES detectors were intentionally operated at a temperature above the superconducting transition where they can absorb higher optical power but have lower responsivity. Operating in this regime allowed the detectors to measure the FTS interferogram signal, but also caused some detectors to respond non-linearly to the signal near the maximum peak of the interferogram. We show that this non-linear response introduces negligible systematics in the measured spectra within the main band as will be discussed later.

The design of the optical coupling between the FTS and the receiver makes it such that, when the two are optimally aligned,
no or minimal sky signal is seen by the targeted detectors.  Additionally, the signal from the FTS source is close to an order of magnitude stronger than the sky signal.  Therefore in the case where stray light enters the system through reflections or insufficient optical coupling, the signal seen by the detector is still dominated by the FTS source as desired.  The continuous modulation allows for further attenuation of any stray unmodulated signal as well.  Because of this, it was possible to take in-field FTS measurements even during times of high atmospheric loading as are typical at the time of year the measurements were performed.

\begin{figure*}[htp]
\centering
\includegraphics[width=.5\textwidth]{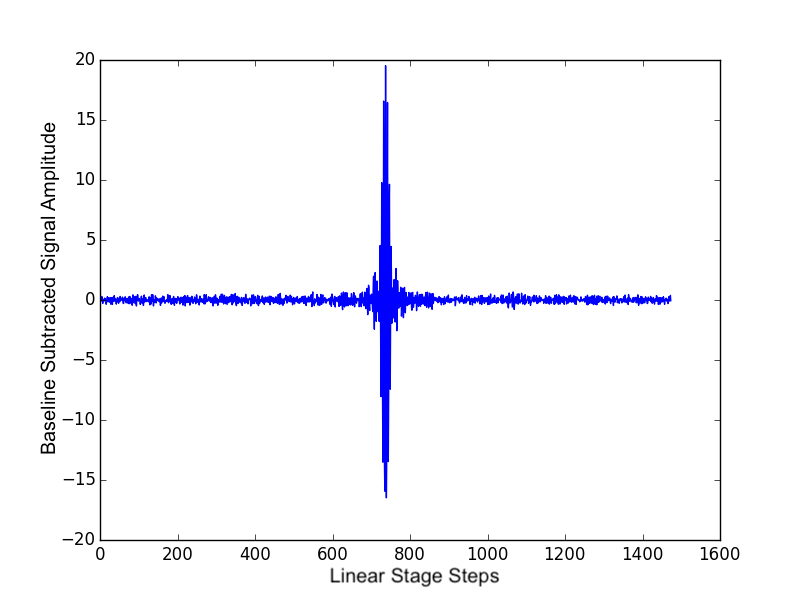}\hfill
\includegraphics[width=.5\textwidth]{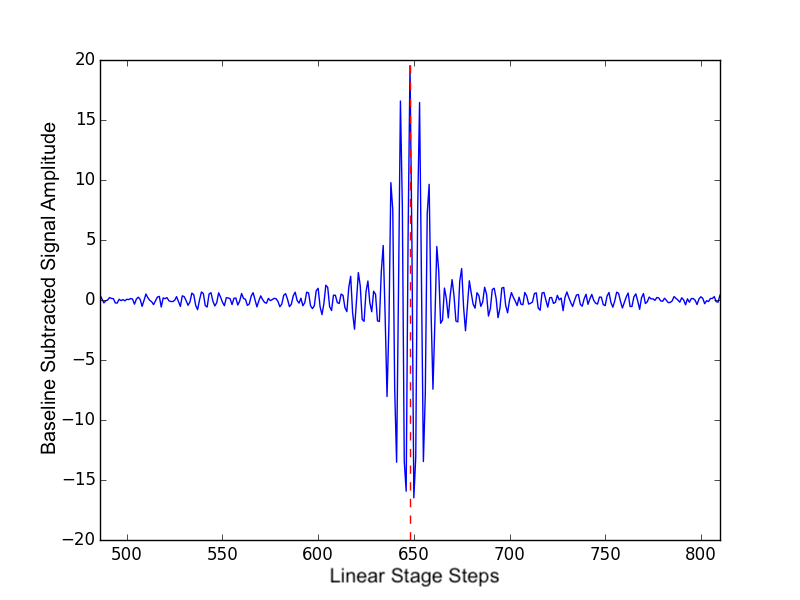}\hfill
\caption[Measured interferogram]{\label{FIG_interferogram} An example of a measured interferogram for one of the \pb{} detectors is shown. The full interferogram (left) and the interferogram near the peak (right) is shown. The red dotted line represents the interferogram peak.}
\end{figure*}

In order to obtain sufficient SNR in the measurements, the PB-FTS movable mirror was stepped in 1400 steps across 30 cm of travel.  Data was recorded at each step with an integration time of 2 seconds, and each FTS run took ${\sim}$90 minutes. An example of a measured interferogram is shown in Figure \ref{FIG_interferogram}.
The $2f$ interferogram signal, given by equation \ref{eqn_2fterm}, in each integration step was demodulated based on the monitored rotation frequency of the rotating output polarizing grid. The demodulated $2f$ signal from each integration step is one data point in the measured interferogram.
In order to remove the effects from gain drifts in each run, a fifth order polynomial was fit and subtracted from the baseline of each double-sided interferogram.

\begin{figure*}[htp]
\centering
\begin{tabular}{cc}
\raisebox{-0.5\height}{\includegraphics[width=.35\textwidth]{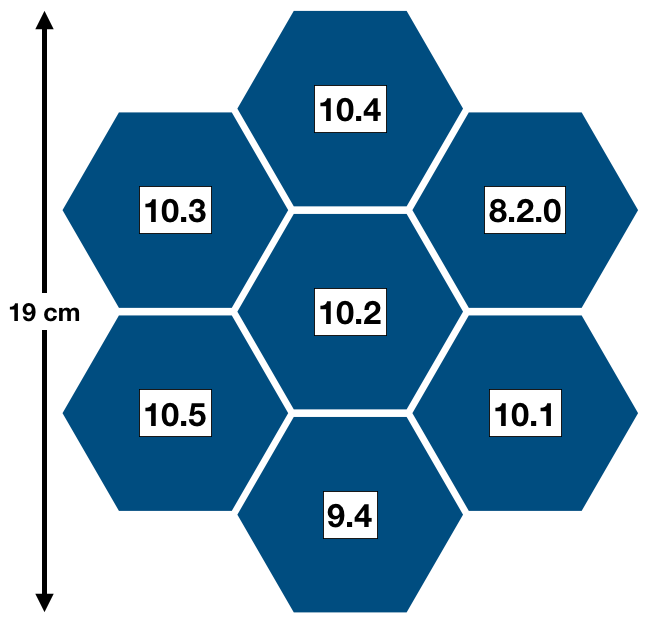}}
\raisebox{-0.5\height}{\includegraphics[width=.65\textwidth]{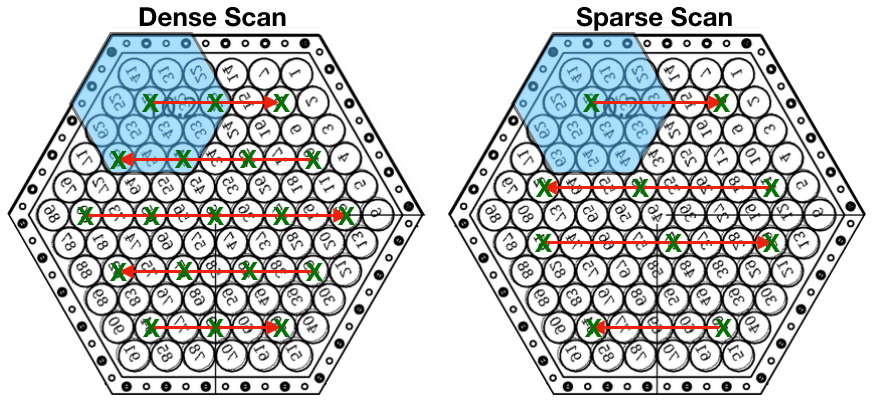}}
\end{tabular}
\caption[FTS scan strategy]{\label{FIG_ftsscan} Image of the \pb{} focal plane with wafer identifications labeled is shown on the left. There are a total of 7 wafers. The dense (center) and sparse (right) scan strategies are illustrated. The green Xs represent the targeted pixels and the red arrow represents the scan order. The PB-FTS can obtain data for the 19 pixels centered around each targeted pixel per run represented by the blue hexagon. The dense scan (19 runs) was used for the central wafer 10.2 and the sparse scan (10 runs) was used for the other 6 wafers. Each run aims at a different set of pixels within the wafer.}
\end{figure*}

The PB-FTS installation, testing, and measurements were done within a month's time. In order to efficiently measure the statistical characteristics of the focal plane detectors, the PB-FTS scanning method was differentiated between detector wafers. The central wafer was scanned most densely with 19 runs. The other 6 wafers were scanned only with 10 runs each. During each run the FTS was aimed at a different set of pixels within the wafer. The scan strategies are illustrated in Figure \ref{FIG_ftsscan}. 
With this scan strategy, theoretically, all the detectors across the focal plane can be measured and repeated measurements can be obtained for a large portion of the detectors in order to achieve higher statistical precision and check for potential systematics effects.

\subsection{Data Analysis}\label{SUBSEC_analysis}

The interferograms taken through the 79 datasets were analyzed, and for detectors that had repeated measurements across datasets the spectra were averaged to improve the SNR. The spectra were obtained with 1 GHz spectral resolution. We obtained detector spectra with SNR $>20$ for 875 detectors ($\sim$69\% of the focal plane): $\sim$82\% of the detectors in the central wafer and $\sim$67\% of the detectors in the other 6 wafers. 
Measured interferograms with very low SNR such that the interferogram peak and central fringes amplitudes were on similar levels to the noise were cut in the data analysis. 
A small number ($<10$ detectors) of extreme outlier spectra in which it was apparent that a spectral band was not being properly measured were manually checked and cut from the final dataset.

In \pb{} the two orthogonal detectors in one pixel pair are labeled ``top'' and ``bottom'' when they are fabricated. For each pixel the top detector's spectrum is peak normalized and then a multiplicative relative gain factor is calculated to normalize its pair's bottom detector spectrum. The relative gain factor is calculated assuming a signal from an atmospheric emission spectrum at 1 mm PWV and elevation $60^{\circ}$ as a typical observing configuration. The gain factor is calculated such that the integrated signal power in the two detectors of a pair are equal. Theoretically this is equivalent in analysis technique to how the relative gain calibration in a pixel would be calibrated if one were to calibrate the detectors using the atmosphere such as using elevation nods. According to the exact calibration methods and sources chosen for CMB data, the emission spectra used in this relative gain factor calculation would need to be adjusted. 

The integrated band center $\nu_{c}$ and integrated bandwidth $\Delta \nu$ are calculated by
\begin{equation}
\nu_{c} = \frac{\int^{\nu_{r}}_{\nu_{l}} \nu S(\nu) d\nu}{\int^{\nu_{r}}_{\nu_{l}} S(\nu) d\nu} ,\ \ \  \text{and} \ \ \ \Delta \nu = \int^{\nu_{r}}_{\nu_{l}} S(\nu) d\nu,
\end{equation}
where $\nu_{l}$ and $\nu_{r}$ are the left (rising) and right (setting) edges of the band respectively. $S(\nu)$ is the detector spectrum. The rising and setting edges of the band are the values for which the normalized spectral response increases above and decreases below a threshold value of 0.050 respectively. This threshold value was chosen to be sufficiently above the statistical error in a single detector's spectrum.

\subsection{Results}\label{SUBSEC_results}

\begin{table*}
\begin{center}
\caption[Detector spectral response statistics]{\label{TABLE_measurements} The measured and calculated detector spectral response statistics are shown. The number of measured detectors, integrated band center, and integrated bandwidth values are given per wafer. The average (AVG) and standard deviation (STD) of the distribution for each wafer are calculated for the band center and bandwidth. The pixel in-band fractional differences are also shown. The pixel in-band fractional difference is calculated as the standard deviation across all in-band data points in the pixel pair difference spectra. The values in the table indicate the average of the distribution for each wafer.}
\begin{tabular}{c|c|c|c|c|c|c}
\hline
\hline
\multirow{2}{*}{\small Wafer} & \multirow{2}{*}{\small \# Detectors} & \multicolumn{2}{c|}{{\small Band Center (GHz)}} & \multicolumn{2}{c|}{{\small{Bandwidth (GHz) \ }}} & {\small Pixel In-band Fractional Diff} \\
\cline{3-7}
&  & {\small{ \ \ AVG \ \ }} & {\small STD} & {\small{ \ \ AVG \ \ }} & {\small STD} & {\small{ \ \ AVG \ \ }}  \\
\hline
8.2.0		& 106	& 136.9 & 0.7 & 30.4 & 1.8 & 0.059 \\
9.4		& 107	& 146.9 & 0.5 & 32.8 & 1.6 & 0.062 \\
10.1		& 113	& 142.1 & 2.5 & 31.8 & 1.8 & 0.062 \\
10.2		& 149	& 143.5 & 0.5 & 32.6 & 1.1 & 0.041 \\
10.3		& 133	& 148.7 & 0.6 & 31.0 & 1.9 & 0.062 \\
10.4		& 154	& 144.0 & 0.5 & 32.2 & 1.2 & 0.044 \\
10.5		& 113	& 145.5 & 0.4 & 31.8 & 1.3 & 0.041 \\
\end{tabular}
\end{center}
\end{table*}

The band center and bandwidth distributions per wafer were calculated and the average and standard deviations are given in Table \ref{TABLE_measurements}.
It was found that the band centers were consistent within a wafer but vary on a per wafer basis by a few GHz. Wafer 8.2.0 was found to have a slightly lower average band center than the rest. Wafer 10.1 was found to have larger variations within a wafer due to a bimodal distribution of the band centers with two peaks at $\sim$140 and $\sim$146 GHz. 
The bandwidths were found to vary by a few GHz within each wafer and between wafers as well. There was no distinct outlier wafer in terms of measured bandwidths.


The in-band fractional difference for each pixel was calculated by taking the difference in spectra (top minus bottom) between detectors in the pixel pair and measuring the standard deviation of the values in the differenced spectra across the band. Table \ref{TABLE_measurements} shows the average value in each wafer's distribution. The left and right edges of the band for each pixel pair were calculated by taking the average left and right edge values of the top and bottom detectors in that pair. 

\begin{figure*}[htp]
\centering
\includegraphics[width=.5\textwidth]{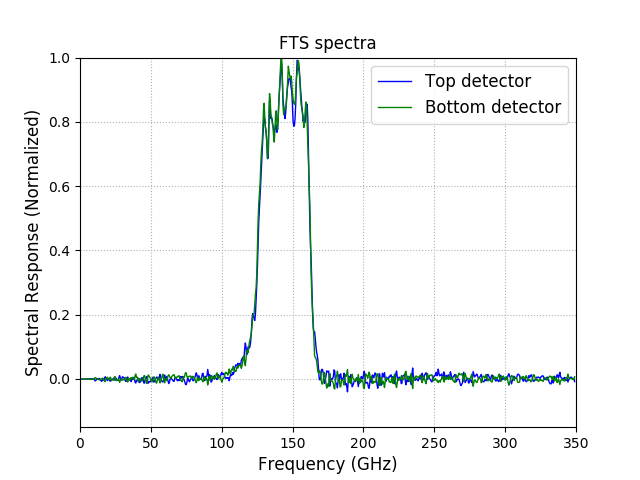}\hfill
\includegraphics[width=.5\textwidth]{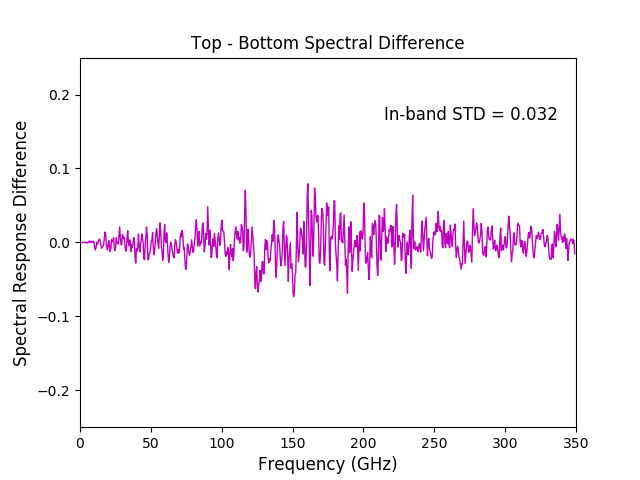}\hfill
\caption[Measured pixel pair spectral difference]{\label{FIG_specdiff} An example of the spectra for one pixel on wafer 10.2 is shown on the left. The spectra for the top and bottom detectors are over-plotted. The differenced spectra is shown on the right. The in-band fractional difference for this pixel is $0.032$. }
\end{figure*}

\begin{figure}[htp]
\centering
\includegraphics[width=.5\textwidth]{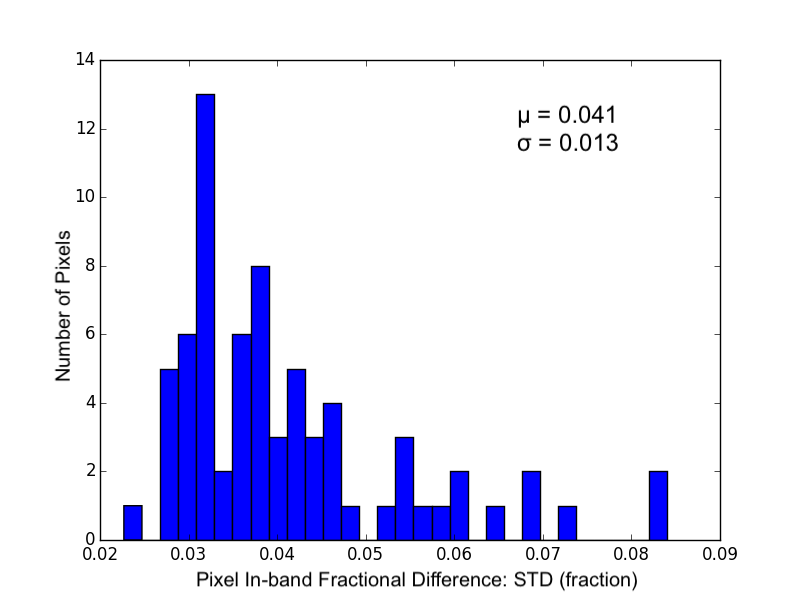}\hfill
\caption[Measured pixel pair spectral difference distributions]{\label{FIG_specdiffdist} The distribution of the pair in-band fractional difference is shown for the central wafer 10.2. The average $\mu$ and spread $\sigma$ of the distribution is shown as well. The distribution shapes are similar across all focal plane wafers, and a sparse tail is typically observed in the distribution.}
\end{figure}

The pixel in-band fractional differences were found to range between 2--17\% across the entire focal plane with averages $<7$\% for each wafer. Example pixel pair spectra and their difference are shown in Figure \ref{FIG_specdiff}. The distribution of the pixel in-band fractional differences for the central wafer 10.2 is shown in Figure \ref{FIG_specdiffdist}. The distribution shapes are similar across all focal plane wafers. Typically the peak of the distribution is located below the average value with a sparse tail toward higher fractional difference values.

\begin{figure*}[htp]
\centering
\includegraphics[width=.5\textwidth]{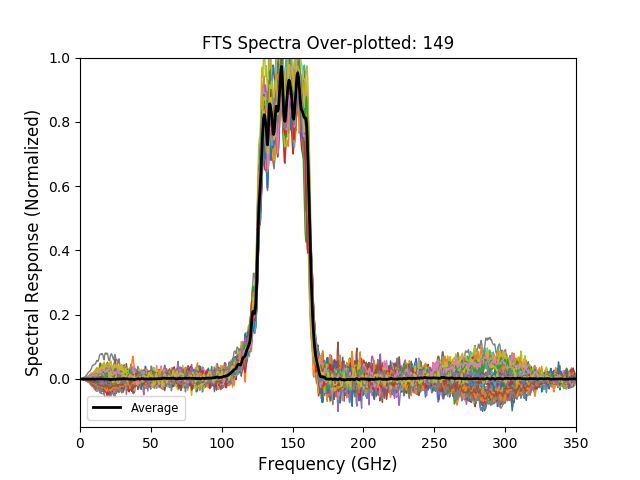}\hfill
\includegraphics[width=.5\textwidth]{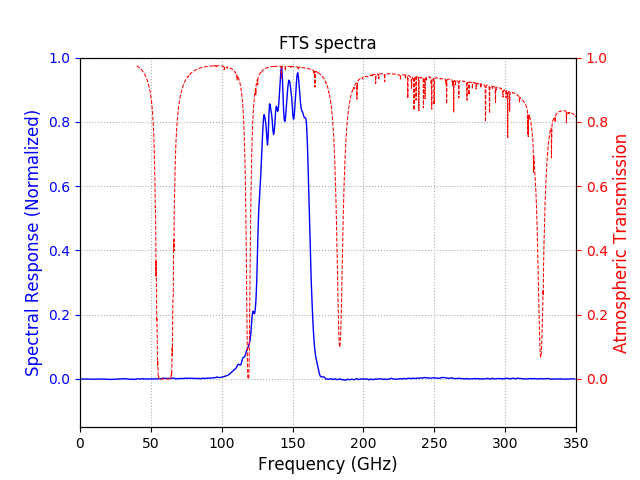}\hfill
\caption[Wafer 10.2 averaged spectrum]{\label{FIG_waferspec1} All the spectra for wafer 10.2 over-plotted are shown (left). The averaged spectrum is over-plotted as a thick black line. The wafer-averaged spectrum for wafer 10.2 is shown (right). The left and right edges of the band are at 114.3 and 167.4 GHz respectively.
The red dashed line indicates the calculated atmospheric transmission at the \pb{} site assuming a PWV level of 1 mm and an elevation angle of $60^{\circ}$.}
\end{figure*}

\begin{figure*}[htp]
\centering
\includegraphics[width=.5\textwidth]{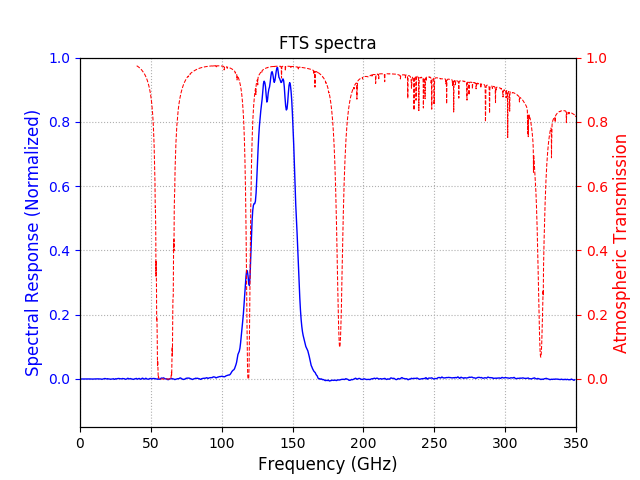}\hfill
\includegraphics[width=.5\textwidth]{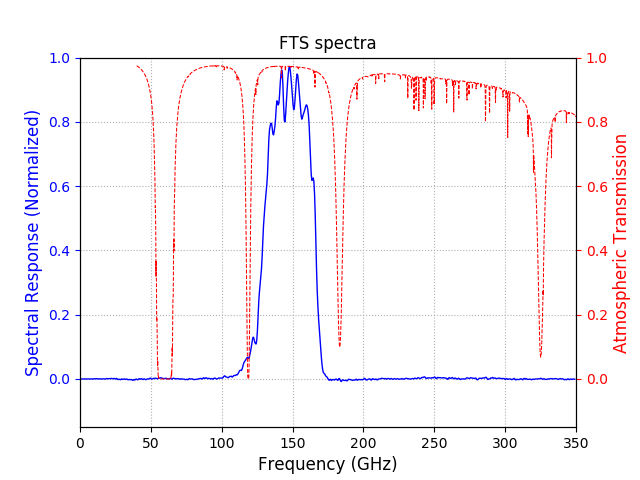}\hfill
\caption[Wafer 8.2.0 and 9.4 averaged spectrum]{\label{FIG_waferspec57} The wafer-averaged spectrum for wafer 8.2.0 (left) and 9.4 (right) are shown. The left and right edges of the band for wafer 8.2.0 are at 110.5 and 162.7 GHz respectively, and for wafer 9.4 are at 115.7 and 170.7 GHz respectively.}
\end{figure*}

\begin{figure*}[htp]
\centering
\includegraphics[width=.5\textwidth]{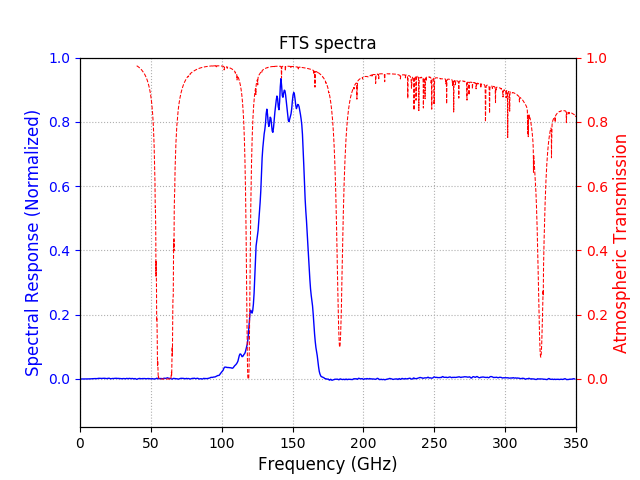}\hfill
\includegraphics[width=.5\textwidth]{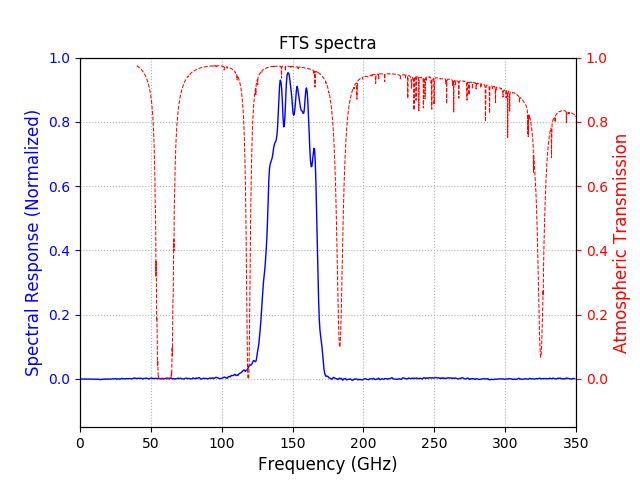}\hfill
\caption[Wafer 10.1 and 10.3 averaged spectrum]{\label{FIG_waferspec36} The wafer-averaged spectrum for wafer 10.1 (left) and 10.3 (right) are shown. The left and right edges of the band for wafer 10.1 are at 111.0 and 168.4 GHz respectively, and for wafer 10.3 are at 121.9 and 172.2 GHz respectively.}
\end{figure*}

\begin{figure*}[htp]
\centering
\includegraphics[width=.5\textwidth]{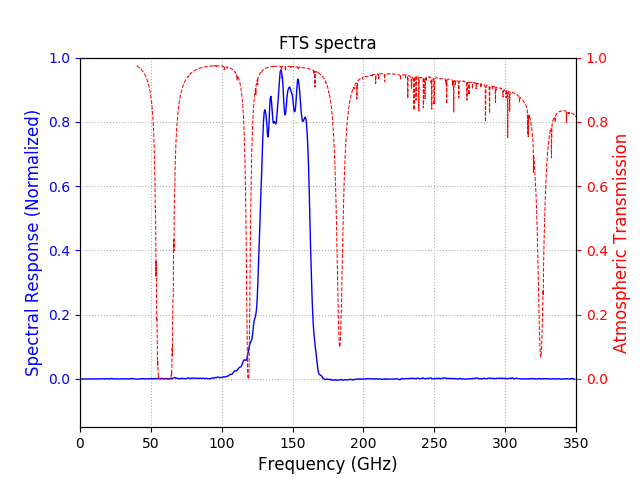}\hfill
\includegraphics[width=.5\textwidth]{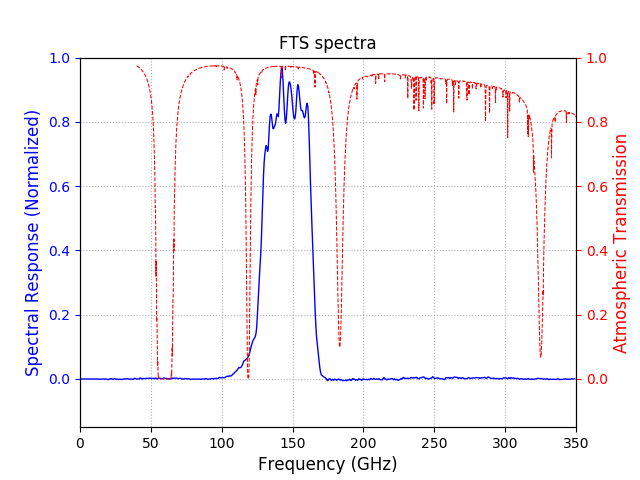}\hfill
\caption[Wafer 10.4 and 10.5 averaged spectrum]{\label{FIG_waferspec24} The wafer-averaged spectrum for wafer 10.4 (left) and 10.5 (right) are shown. The left and right edges of the band for wafer 10.4 are at 115.2 and 167.4 GHz respectively, and for wafer 10.5 are at 115.2 and 169.3 GHz respectively.}
\end{figure*}

The averaged spectra per wafer were also calculated. These are shown in Figures \ref{FIG_waferspec1}, \ref{FIG_waferspec57}, \ref{FIG_waferspec36}, and \ref{FIG_waferspec24}. It can be seen in Figure \ref{FIG_waferspec1} that some detectors show a ``bump'' in the spectrum at $\sim$300 GHz and at low frequencies below $\sim$30 GHz. This effect is hypothesized to be due to the non-linear response when high power is incident on the detector. This non-linear response was found to become more apparent with more incident power. It is expected that due to the baseline polynomial subtraction in the interferogram, spectral power below $\sim$5 GHz has been attenuated.

These bumps in the spectra were only apparent in detectors in which the polarization-sensitive axis was more co-aligned with the PB-FTS beam-splitter orientation, a situation which creates more incident power on the detectors. 
Each wafer is designed to contain two types of pixels: one type in which the pixel's polarization-sensitive axes are oriented at 0 and 90 degrees, and another type in which the pixel's polarization-sensitive axes are oriented at 45 and 135 degrees. Hence within a wafer, at maximum only half of the detectors in that wafer can potentially show this non-linear effect at all.
On top of that, because there are three different overall wafer rotation orientations and thus three different wafer polarization configurations in the \pb{} focal plane, this non-linear effect was apparent mostly in specific wafers that had the same wafer orientation, not in all wafers.

These spectral bumps were predominantly observed in the detector spectra from parts of the central wafer 10.2 that also had the highest optical coupling efficiency to the PB-FTS. This effect was suppressed or not observed in the detector spectra when coupling the PB-FTS to detectors located further away from the focal plane center due to the decreased coupling efficiency as explained in Section \ref{SUBSEC_mirrors}. Even within the central wafer, rows of pixels of the same polarization orientation pixel type would show this non-linear effect while neighboring pixel rows of the other polarization orientation pixel type did not show this effect. 

As can be seen from Table \ref{TABLE_measurements}, comparing the standard deviations in the calculated band centers and bandwidths for each wafer, no noticeable or outstanding statistical differences were observed between the central wafer where detectors often exhibited these spectral bumps and the edge wafers that mostly did not. 
Because the central wafer contains a comparable fraction of detectors that showed this effect to those that did not, one could expect that the standard deviation of calculated band centers and bandwidths would noticeably differ (become potentially larger) compared to the standard deviations of the other wafers that did not show this effect. 
Because this does not appear to be the case, it is expected that this non-linear effect has a sub-dominant impact on the calculated band center and bandwidth compared to the measurement errors and intrinsic detector differences within each wafer. 

\subsection{Statistical and Systematic Errors}\label{SUBSEC_errors}

The statistical and systematic errors per wafer are summarized in Table \ref{TABLE_errors}. All errors are fractional errors in each spectral data point relative to the peak value of each spectrum that is normalized as explained previously. 
\begin{table}
\begin{center}
\caption[Detector spectral response fractional errors]{\label{TABLE_errors} The estimated statistical and systematic fractional errors in the spectral measurements for each wafer are shown. The average of the statistical error value distribution for each wafer is shown. The average systematic error value for the subset of detectors that were measured multiple times over different runs per wafer is also shown. These are measured fractional errors in each frequency data point of the spectra with the normalization as explained previously.}
\begin{tabular}{c|c|c|c}
\hline
\hline
Wafer & \# Detectors & Statistical Error & Systematic Error \\
\hline
8.2.0		& 106	& 0.018 & 0.033 \\
9.4		& 107	& 0.019 & 0.048 \\
10.1		& 113	& 0.020 & 0.045 \\
10.2		& 149	& 0.017 & 0.041 \\
10.3		& 133	& 0.020 & 0.049 \\
10.4		& 154	& 0.016 & 0.034 \\
10.5		& 113	& 0.021 & 0.035 \\
\end{tabular}
\end{center}
\end{table}
The statistical error is the RMS noise measured outside of the detector band in each individual detector.
The final spectrum for each detector across the focal plane was measured to a statistical precision of 0.7--4.4\%. The seven wafer-averaged spectra all have statistical errors of $<0.4$\%. 


The systematic error includes uncertainties arising from both time-variability in the measurement and from errors in alignment of the PB-FTS with the \pb{} instrument.  The combined errors introduced by these two effects were estimated by comparing and calculating the variation in the spectra for detectors that were measured multiple times over different runs and thus with differing PB-FTS alignments relative to the HTT.  These systematic errors were calculated from the standard deviation across all frequencies when differencing the compared spectra and were found to be 3.3--4.9\% depending on the wafer. Of this, the time-variability error, measured separately over a timespan of 6 hours with the same alignment, was estimated to be $<2.3$\%.

Wafer 8.2.0 is expected to have the least sensitivity due to the lower band center causing the signal to have larger contamination from the atmospheric oxygen line emission which may also introduce larger temperature-to-polarization systematic leakage. 
Wafer 8.2.0 was known from the detector fabrication phase to be an outlier compared to the rest of the focal plane wafers.

\section{Discussion and Future Applications}
\label{SEC_future}

The high throughput and specifically designed output parabolic mirror provided high efficiency optical coupling across the detector focal plane of the \pb{} receiver. The continuously rotating output polarizer provided efficient signal modulation that allowed for measuring the instrument and detector spectra with high SNR. The PB-FTS functioned as designed and successfully measured high precision instrument and detector spectra for ${\sim}$69\% of the detector array directly in the field.

The PB-FTS method of optical coupling is different from the methods used by other CMB experiments. Keck Array and BICEP3 couple their FTS on the sky side of all their telescope optics where the rays are already collimated\cite{BK2014FTS}. This method of coupling to the collimated rays has the advantages that the optical coupling design is very simple and that the spectral response of all optical elements in the telescope can be characterized at once. The disadvantage, however, is that in order to fill the entire beam of the detectors, the FTS main beam aperture size must be equivalent to or larger than that of the main aperture of the telescope. A higher throughput FTS with larger coupling optics would be required. Because of this, this method can only be reasonably applied to small aperture telescopes, and becomes challenging and cost ineffective for larger aperture telescopes like \pb{}.  Even with small aperture refractive telescopes, next generation experiments plan to increase main aperture size to scales of $\sim$50 cm\cite{SOForecast2019,BICEP32018}, which will be difficult to couple to efficiently with this technique. 

ACT couples its FTS in front of their receiver using refractive coupling lenses\cite{Datta2016}. This method of coupling using lenses is a flexible technique that can also allow for low aberration optical coupling according to the number of lenses and the lens shapes. This technique allows for a larger range of freedom in positioning the FTS, and typically refractive coupling optics occupy less volume compared to reflective coupling optics. A disadvantage is that refractive coupling lenses require anti-reflection (AR) coatings in order to sustain high transmission across various frequency ranges. Typically different sets of optimized AR coatings are used to cover the different observation bands of CMB experiments. Even though the Martin-Puplett interferometer itself can be used across a very wide range of frequencies, the coupling lenses would need to be altered according to the frequency band of interest. With next generation CMB experiments extending to cover ever wider ranges of observation frequencies\cite{SOForecast2019}, this method would require multiple refractive coupling lenses to sustain high efficiency optical coupling across all observation frequencies. 

The strength of the PB-FTS coupling method is that it provides effective beam-filling and low-aberration optical coupling while keeping the FTS size reasonable and having no frequency dependence in the coupling optics by using mirrors only. 
The only potential disadvantage of this technique is that the coupling mirrors produce an innately off-axis optical system, which may not be suitable for some telescope designs that have limited space. For \pb{}, which employs off-axis optics to begin with, the spatial concern is not an issue. 
This custom parabolic mirror coupling technique is theoretically applicable to various off-axis Gregorian Dragone-type telescopes, and a similar theoretical concept can be applied for other Dragone-type telescopes as well.

This PB-FTS was designed to also be compatible with \pb{}-2, the next-generation receiver of \pb{}, in the field. Three \pb{}-2 receivers, each mounted on a separate telescope, will comprise the Simons Array (SA).  The three-telescope array will operate over a total of four different frequency bands centered at 90, 150, 220, and 270 GHz, with a total of 22,764 detectors\cite{Westbrook2018}. The PB-FTS will be used to characterize the spectral response of all the SA receivers. 

SA uses the same HTT design to couple to receivers with re-designed higher-throughput optics.
The PB-FTS optical design is driven by the primary and secondary reflectors, and hence the PB-FTS is compatible with SA as well. As described in section \ref{SUBSEC_mirrors}, the optical coupling efficiency decreases as a function of increased distance from the optical central axis. For the larger focal plane of the SA receivers this effect is larger. However, Zemax optical ray tracing simulations indicate that greater than 76\% of the equally spaced geometric rays that pass through the SA receiver Lyot stop are expected to be unvignetted and reach the PB-FTS source even when coupling the PB-FTS to the outermost SA receiver pixels. Even though the optical coupling efficiency decreases for edge regions, the PB-FTS has enough throughput such that it will still couple well across the SA focal plane and is compatible with the SA receivers. 

Another characteristic of the PB-FTS that has merits for SA is the continuous modulation by the output polarizing grid. As explained in Section \ref{SUBSEC_mod}, the $4f$ component interferogram is dependent on the relative angle between the output polarizer and detector polarization axis. SA uses very broadband sinuous antennas that are known to have a polarization angle wobble \cite{saini1996}. 
Hence the $4f$ component potentially contains information about the polarization-sensitivity as a function of frequency across the observation band.
With further analytical study, the spectral response of the SA detectors as a function of the polarization angle may possibly be extracted from the $4f$ component. 

\section{Conclusions}
\label{SEC_conclusions}

The PB-FTS has two unique features compared to other FTS calibrators used in the CMB field: high throughput and a continuously rotating output polarizer. The PB-FTS also has a specifically design output parabolic coupling mirror for optimal coupling to the \pb{} receiver. These elements enabled efficient data taking for in-field spectral characterization of the \pb{} receiver performed in April 2014.  After data cuts to eliminate detectors whose interferograms had insufficient signal-to-noise, spectra with a resolution of 1 GHz were obtained for 875 \pb{} detectors, corresponding to $\sim$69\% of the focal plane.  Analysis of these spectra quantified the spectral in-band difference between pixel pairs that can introduce temperature-to-polarization systematic errors.

The successful FTS run proved that the PB-FTS is capable of measuring the spectral response of the \pb{} instrument with high precision.  For the PB-FTS specifically, future plans include in-field spectral measurements of the SA receivers following their deployment, but the techniques and design choices employed by this FTS can be applied more broadly for use with other instruments as well.

\begin{acknowledgements}
We dedicate this paper to Professor Hans Paar. 
FM acknowledges the support by the World Premier International Research Center Initiative (WPI), MEXT, Japan and JSPS fellowship (Grant number JP17F17025).
MA acknowledges support from CONICYT UC Berkeley-Chile Seed Grant (CLAS fund) Number 77047, Fondecyt project 1130777 and 1171811, DFI postgraduate scholarship program and DFI Postgraduate Competitive Fund for Support in the Attendance to Scientific Events.
FB acknowledges support from an Australian Research Council Future Fellowship (FT150100074).
YC acknowledges the World Premier International Research Center Initiative (WPI), MEXT, Japan and support from the JSPS KAKENHI grant Nos. 18K13558 and 18H04347.
GF is supported by the European Research Council under the European Union's Seventh Framework Programme (FP/2007-2013) / ERC Grant Agreement No. 616170.
Kavli IPMU was supported by World Premier International Research Center Initiative (WPI), MEXT, Japan. This work was supported by the JSPS Core-to-Core Program.
Support from the Ax Center for Experimental Cosmology is gratefully acknowledged. 
\end{acknowledgements}

\appendix*
\section{Modulation Theory Derivation}
\label{APP_modulation}

The derivation of the signal measured by a polarization-sensitive detector coupled to a Martin-Puplett interferometer uses the Jones formalism and definitions from Martin\cite{Martin1982} as the starting point. Throughout the derivation, ideal polarizers and perfect alignment within the FTS has been assumed for simplicity.
The signal amplitude through the interferometer output port $A$ coupled to a polarization-sensitive detector is given by:
\begin{eqnarray}
E_{\rm det} & = & R_{\rm det}(\alpha-\theta(t)) A(\theta(t)) \nonumber \\
& = & R_{\rm det}(\alpha-\theta(t)) T_{\rm OP} R_{\rm OP}(\theta(t)) D E_{\rm IP}.
\end{eqnarray}
The signal from the input polarizer $E_{\rm IP}$ is
\begin{equation}
E_{\rm IP} = \begin{bmatrix}
	S_{i} \\
	P_{i} \\
	\end{bmatrix}
\end{equation}
where $S_{i}$ and $P_{i}$ are the signal amplitude components going into the beam-splitter from the input polarizer. These relate to the signals $I$ and $J$ from the interferometer input ports according to equations 40 and 41 in Martin\cite{Martin1982}:
\begin{equation}
\begin{bmatrix}
S_{i} \\
P_{i} \\
\end{bmatrix} 
= 
\begin{bmatrix}
1 & 0 \\
0 & 0 \\
\end{bmatrix}
\begin{bmatrix}
I_{s} \\
I_{p} \\
\end{bmatrix}
+
\begin{bmatrix}
0 & 0 \\
0 & 1 \\
\end{bmatrix}
\begin{bmatrix}
J_{s} \\
J_{p} \\
\end{bmatrix}
\end{equation}
where ideal polarizers have been assumed. Assuming perfect alignment, the relations between $S_{i}$, $P_{i}$, $I_{s}$, and $J_{p}$ are
\begin{eqnarray}
S_{i}^{2} + P_{i}^{2} & = & I_{s}^{2} + J_{p}^{2} \label{eqn_spij1} \\
S_{i}^{2} - P_{i}^{2} & = & I_{s}^{2} - J_{p}^{2} \label{eqn_spij2} \\
S_{i} P_{i}^{\ast} - S_{i}^{\ast} P_{i} & = & 0 \label{eqn_spij3} \\
S_{i} P_{i}^{\ast} + S_{i}^{\ast} P_{i} & = & 0 . \label{eqn_spij4}
\end{eqnarray}
The phase shift matrix $D$ is given by equation 24 in Martin\cite{Martin1982}:
\begin{equation}
D = \frac{1}{2} \begin{bmatrix}
	d_{-}  & -d_{+} \\
	d_{+} & -d_{-} \\
	\end{bmatrix} .
\end{equation}
Here it has been assumed that the beam-splitter efficiency is ideal. Assuming perfect alignment, the relation between $d_{+}$, $d_{-}$, $d_{f}$, and $d_{m}$ are given by equation 29 in Martin\cite{Martin1982}:
\begin{eqnarray}
d_{+}^{2} + d_{-}^{2} & = & 2 \left( d_{f}^{2} + d_{m}^{2} \right) \label{eqn_d1} \\
d_{+}^{2} - d_{-}^{2} & = & 4 \cos \left( \Delta \right) \label{eqn_d2} \\
d_{+} d_{-}^{\ast} + d_{+}^{\ast} d_{-} & = & 2 \left( d_{f}^{2} - d_{m}^{2} \right) \label{eqn_d3} \\
d_{+} d_{-}^{\ast} - d_{+}^{\ast} d_{-} & = & 4 i \sin \left( \Delta \right) . \label{eqn_d4}
\end{eqnarray}
The output polarizer matrix $R_{\rm OP}$ is a standard rotation matrix.
\begin{equation}
R_{\rm OP} = \begin{bmatrix}
	\cos \left( \theta(t) \right) & \sin \left( \theta(t) \right) \\
	-\sin \left( \theta(t) \right) & \cos \left( \theta(t) \right) \\
	\end{bmatrix} .
\end{equation}
The transmission matrix through the output polarizer $T_{\rm OP}$ is given by:
\begin{equation}
T_{\rm OP} = \begin{bmatrix}
	0 & 0 \\
	0 & 1 \\
	\end{bmatrix} .
\end{equation}
The polarized detector matrix $R_{\rm det}$ is also a standard rotation matrix.
\begin{equation}
R_{\rm det} = \begin{bmatrix}
	\cos \left( \alpha-\theta(t) \right) & \sin \left( \alpha-\theta(t) \right) \\
	-\sin \left( \alpha-\theta(t) \right) & \cos \left( \alpha-\theta(t) \right) \\
	\end{bmatrix} .
\end{equation}
Now solving for $E_{\rm det}$
\begin{eqnarray}
E_{\rm det} & = & \begin{bmatrix}
	E_{\mathrm{det},r} \\
	E_{\mathrm{det},t} \\
	\end{bmatrix} \nonumber \\
& = & \frac{1}{2} \footnotesize{\begin{bmatrix}
	\bigl( \left( d_{+} P_{i} - d_{-} S_{i} \right) \sin \left( \theta(t) \right) \\
	+ \left( d_{+} S_{i} - d_{-} P_{i} \right) \cos \left( \theta(t) \right) \bigr) \sin \left( \alpha-\theta(t) \right) \\
	\bigl( \left( d_{+} P_{i} - d_{-} S_{i} \right) \sin \left( \theta(t) \right) \\
	+ \left( d_{+} S_{i} - d_{-} P_{i} \right) \cos \left( \theta(t) \right) \bigr) \cos \left( \alpha-\theta(t) \right) \\
	\end{bmatrix}}
\end{eqnarray}
where the subscripts $r$ and $t$ represent the reflected and transmitted components. The transmitted component is the measured detector signal. Therefore the signal power measured by the detector is
\begin{eqnarray}
E_{\mathrm{det},T} E_{\mathrm{det},T}^{\ast} & = & \frac{1}{4} \cos^{2} \left( \alpha-\theta(t) \right) \bigl[ \left( d_{+} P_{i} - d_{-} S_{i} \right) \sin \left( \theta(t) \right) \nonumber \\
& & + \left( d_{+} S_{i} - d_{-} P_{i} \right) \cos \left( \theta(t) \right) \bigr] \nonumber \\
& & \bigl[ \left( d_{+}^{\ast} P_{i}^{\ast} - d_{-}^{\ast} S_{i}^{\ast} \right) \sin \left( \theta(t) \right) \nonumber \\
& & + \left( d_{+}^{\ast} S_{i}^{\ast} - d_{-}^{\ast} P_{i}^{\ast} \right) \cos \left( \theta(t) \right) \bigr] .
\end{eqnarray}
Simplifying the expression and finally substituting equations \ref{eqn_spij1} through \ref{eqn_spij4} and equations \ref{eqn_d1} through \ref{eqn_d4}, one obtains the results
\begin{equation}
E_{\mathrm{det},T} E_{\mathrm{det},T}^{\ast} = E_{\rm DC}^{2} + E_{2f}^{2} + E_{4f}^{2} 
\end{equation}
\begin{eqnarray}
E_{\rm DC}^{2}  & = & a_{1} + \frac{1}{2} a_{2} \cos \left( 2 \alpha \right) +  \frac{1}{2} a_{3} \sin \left( 2 \alpha \right) \nonumber \\
& = & \frac{1}{2} a_{2} (\Delta) \cos \left( 2 \alpha \right) + C_{\rm DC} \\
E_{2f}^{2}  & = & a_{1} \cos \left( 2 \theta(t) - 2 \alpha \right) + a_{2} \cos \left( 2 \theta(t) \right) + a_{3} \sin \left( 2 \theta(t) \right) \nonumber \\
& = & a_{2} (\Delta) \cos \left( 2 \theta(t) \right) + C_{2f} \\
E_{4f}^{2}  & = & \frac{1}{2} a_{2} \cos \left( 4 \theta(t) - 2 \alpha \right) + \frac{1}{2} a_{3} \sin \left( 4 \theta(t) - 2 \alpha \right) \nonumber \\
& = & \frac{1}{2} a_{2} (\Delta) \cos \left( 4 \theta(t) - 2 \alpha \right) + C_{4f} 
\end{eqnarray}
where $a_{i}$ terms are given by
\begin{eqnarray}
a_{1} & = & \frac{1}{8} \left( d_{f}^{2} + d_{m}^{2} \right) \left( I_{s}^{2} + J_{p}^{2} \right) \\
a_{2} & = & \frac{1}{4} \left( I_{s}^{2} - J_{p}^{2} \right) \cos \left( \Delta \right) \\
a_{3} & = & \frac{1}{8} \left( d_{f}^{2} - d_{m}^{2} \right) \left( I_{s}^{2} + J_{p}^{2} \right) .
\end{eqnarray}

\bibliography{ms}

\end{document}